\documentclass[11pt]{article}
\usepackage{amsfonts}
\usepackage{amsmath}
\newcommand{\ud}{\mathrm{d}}
\begin{document}
\begin{titlepage}
\begin{flushright}
CP3-06-05\\
ICMPA-MPA/2006/35\\
April 2007\\
\end{flushright}
\begin{centering}
 
{\ }\vspace{1cm}
 
{\Large\bf Gauge Invariant Factorisation and Canonical Quantisation of}

\vspace{5pt}

{\Large\bf Topologically Massive Gauge Theories in Any Dimension}

\vspace{0.7cm}

Bruno Bertrand

\vspace{0.3cm}

{\em Center for Particle Physics and Phenomenology (CP3)}\\
{\em Institut de Physique Nucl\'eaire}\\
{\em D\'epartement de Physique, Universit\'e catholique de Louvain (U.C.L.)}\\
{\em 2, Chemin du Cyclotron, B-1348 Louvain-la-Neuve, Belgium}\\
{\em E-mail: {\tt Bruno.Bertrand@fynu.ucl.ac.be}}

\vspace{0.8cm}

Jan Govaerts\footnote{On sabbatical
leave from the Center for Particle Physics and Phenomenology (CP3),
Institut de Physique Nucl\'eaire, Universit\'e catholique de Louvain (U.C.L.),
2, Chemin du Cyclotron, B-1348 Louvain-la-Neuve, Belgium,
E-mail: {\tt Jan.Govaerts@fynu.ucl.ac.be}.}

\vspace{0.3cm}

{\em Institute of Theoretical Physics}\\
{\em Department of Physics, University of Stellenbosch}\\
{\em Stellenbosch 7600, Republic of South Africa}\\

\vspace{0.3cm}

{\em UNESCO International Chair in Mathematical Physics and 
Applications (ICMPA)}\\
{\em University of Abomey-Calavi}\\
{\em 072 B.P. 50, Cotonou, Republic of Benin}\\

\vspace{0.7cm}

\begin{abstract}
\noindent
Abelian topologically massive gauge theories (TMGT) provide a topological mechanism to generate mass for a bosonic $p$-tensor field in any spacetime dimension. These theories include the 2+1 dimensional Maxwell-Chern-Simons and 3+1 dimensional Cremmer-Scherk actions as particular cases. Within the Hamiltonian formulation, the embedded topological field theory (TFT) sector related to the topological mass term is not manifest in the original phase space. However through an appropriate canonical transformation, a gauge invariant factorisation of phase space into two orthogonal sectors is feasible. The first of these sectors includes canonically conjugate gauge invariant variables with free massive excitations. The second sector, which decouples from the total Hamiltonian, is equivalent to the phase space description of the associated non dynamical pure TFT. Within canonical quantisation, a likewise factorisation of quantum states thus arises for the full spectrum of TMGT in any dimension. This new factorisation scheme also enables a definition of the usual projection from TMGT onto topological quantum field theories in a most natural and transparent way. None of these results rely on any gauge fixing procedure whatsoever.

\end{abstract}

\vspace{10pt}

\end{centering} 

\vspace{125pt}

\end{titlepage}

\setcounter{footnote}{0}

\section{Introduction}
\label{sec:Intro}

Topological field theories (TFT, see \cite{Birmingham:1991ty} for a review) have played an important role in a wide range of fields in mathematics and physics ever since they were first constructed by A.~S.~Schwarz \cite{Schwarz:TFT} and E.~Witten \cite{Witten:1982im}. These theories actually possess so large a gauge freedom that their physical, namely their gauge invariant observables solely depend on the topology (more precisely, the diffeomorphism equivalence class) of the underlying manifold. Another related feature of TFT is the absence of propagating physical degrees of freedom. Upon quantisation, these specific properties survive, possibly modulo some global aspects related to quantum anomalies. As a consequence, topological quantum field theories (TQFT) often have a finite dimensional Hilbert space and are quite generally solvable, even though their formulation requires an infinite number of degrees of freedom. There exists a famous classification scheme for TQFT, according to whether they are of the Schwarz or of the Witten type \cite{Birmingham:1991ty}.

As a class of great interest, TFT of the Schwarz type have a classical action which is explicitly independent of any metric structure on the underlying manifold and does not reduce to a total divergence or surface term. The present work focuses on all such theories defined by a sequence of abelian $B\wedge F$ theories for manifolds $\mathcal{M}$ of any dimension $(d+1)$ \cite{Schwarz:TFT,Horowitz:1989ng,Blau:1989bq}. Given a real valued $p$-form field $A$ in $\Omega^p(\mathcal{M})$ and a real valued $(d-p)$-form field $B$ in $\Omega^{d-p}(\mathcal{M})$, the general TFT action of interest is of the form
\begin{equation} \label{def:BF_Action}
S_{B\wedge F}[A,B] = \kappa\, \int_{\mathcal{M}} (1-\xi)\, F\wedge B - (-1)^p\, \xi\, A\wedge H  ,
\end{equation}
$\kappa$ being some real normalisation parameter of which the properties are specified throughout the discussion hereafter.
This action is invariant under two independent classes of finite abelian gauge transformations acting separately in either the $A$-
or $B$-sector,
\begin{equation} \label{def:BF_gauge}
A'= A + \alpha , \qquad  B' =  B + \beta ,
\end{equation}
where $\alpha$ and $\beta$ are closed $p$- and $(d-p)$-forms on $\mathcal{M}$, respectively. The derived quantities $F=\ud A$ and $H=\ud B$ are the gauge invariant field strengths associated to $A$ and $B$. The arbitrary real variable $\xi$ introduced in order to parametrise any possible surface term is physically irrelevant for an appropriate choice of boundary conditions on $\mathcal{M}$. Given the definition of the wedge product, $\wedge$, the integrand in (\ref{def:BF_Action}) is a ($d+1$)-form, the integration of which over $\mathcal{M}$ does not require a metric. In the particular situation when the number of spatial dimensions $d$ is even and such that $d=2p$ with $p$ itself being odd, in addition to the $B\wedge F$ theories defined by (\ref{def:BF_Action}) there exist TFT of the Schwarz type involving only the single $p$-form field $A$ with the following action\footnote{If $p$ is even with $d=2p$, this action reduces to a surface term.},
\begin{equation} \label{def:AF_Action}
S_{A\wedge F}[A] = \kappa \, \int_{\mathcal{M}}  A\wedge F  .
\end{equation}
These theories are said to be of the $A\wedge F$ type. They include the abelian Chern-Simons theory in 2+1 dimensions \cite{Schwarz:TFT,Witten:1988hf}.

This sequence of TFT of the Schwarz type formulated in any dimension, and related to one another through dimensional reduction \cite{Barcelos-Neto:2001dp}, possesses some fascinating properties. First, the space of gauge inequivalent classical solutions is isomorphic to $H^p(\mathcal{M}) \times H^{d-p}(\mathcal{M})$, $H^p(\mathcal{M})$ being the $p^{\textrm{th}}$ cohomology group of the manifold $\mathcal{M}$. Second, the types of topological terms contributing to these actions define generalisations to arbitrary dimensions of ordinary two-dimensional anyons. Namely, non local holonomy effects give rise to exotic statistics for the extended objects which may be coupled to the higher order tensor fields \cite{Fustero-Harvey,Baez:2006un}. Third, these types of quantum field theories display profound connections between mathematics and physics for what concerns topological properties related, say, to the motion group, the Ray-Singer torsion and link theory. These connections appear within the canonical quantisation\footnote{When $\mathcal{M}=\mathbb{R}\times \Sigma$, the physical Hilbert space is the set of square integrable functions on $H^p(\Sigma)$ \cite{Horowitz:1989ng}.} of these systems \cite{Szabo:1999gm,Bergeron:1994ym}.

Furthermore, within the context of dynamical relativistic (quantum) field theories in any spacetime dimension, which is a general framework of potential relevance to fundamental particle physics as well as mathematical investigations for their own sake, such topological $B\wedge F$ terms may be considered to define couplings between two independent tensor fields whose dynamics is characterised by the following action, provided the spacetime manifold $\mathcal{M}$ is endowed now with a Lorentzian metric structure (of mostly negative signature) allowing for the introduction of the associated Hodge $\ast$ operator, 
\begin{eqnarray} \label{def:TMGT_Action}
S_{\textrm{TMGT}}[A,B] & = & \int_{\mathcal{M}} \frac{1}{2 \, e^2} \, (-1)^{p} \, F\wedge\ast F + \frac{1}{2 \, g^2} \, (-1)^{d-p} \, H\wedge\ast H \nonumber\\ 
& & + \kappa \int_{\mathcal{M}} (1-\xi) \, F\wedge B - (-1)^p \, \xi \, A\wedge H  .
\end{eqnarray}
The notations are those introduced previously. Given a choice of units such that $c=1$, the physical dimensions of $A$ and $B$ are $L^{-p}$ and $L^{-d+p}$, respectively, whereas the multiplicative constant $\kappa$ possesses the same physical dimension as the action. The parameters $e$ and $g$ are arbitrary real constants corresponding to coupling constants when matter fields coupled to $A$ and $B$ are introduced. Without loss of generality for the present analysis, the parameters $e$ and $g$ are assumed to be strictly positive. In 3+1 dimensions, one recovers the famous Cremmer-Scherk action \cite{Cremmer-Scherk,Allen:1990gb} and in 2+1 dimensions, the doubled Chern-Simons theory \cite{Jackiw:1997jg}. It is well known that the topological terms generate a mass for the dynamical tensor fields without breaking gauge invariance. Introducing an appropriate choice of gauge fixing, it is possible to render one of the tensor fields massive through a combination with the other tensor field \cite{Allen:1990gb}. In the particular circumstance that $d=2p$ with $p$ odd, a topological term of the $A\wedge F$ type (\ref{def:AF_Action}) generates also a mass even though the action involves a single $p$-form field $A$,
\begin{equation} \label{def:MCS_Action}
S_{\textrm{TMGT}}[A] = \int_{\mathcal{M}} \frac{-1}{2e^2} \, F\wedge\ast F + \frac{\kappa}{2} \, A\wedge F .
\end{equation}
In 2+1 dimensions, this action defines the well-known Maxwell-Chern-Simons theory \cite{MCS}.

The outline of the paper is as follows. Section~\ref{sec:Classical} discusses a new property of the abelian TMGT valid whatever the number of space dimensions and the value of $0\le p\le d$ for the $p$- and $(d-p)$-form fields: the ``Physical-Topological" factorisation of their degrees of freedom. This result is achieved within the Hamiltonian formulation through a canonical transformation of classical phase space leading to two independent and decoupled sectors\footnote{In other words, the Poisson brackets of variables belonging to the two distinct sectors vanish identically.}. The first of these sectors, namely the ``physical" one, consists of gauge invariant variables which are canonically conjugate and describe free massive propagating physical degrees of freedom. The second sector, namely the ``topological" one, consists of canonically conjugate gauge variant variables which are decoupled from the total Hamiltonian and, hence, are non dynamical. This sector is equivalent to a pure TFT of the $A\wedge F$ or $B\wedge F$ type. This factorisation enables the identification of a mass generating mechanism for any $p$-form (or, by dualisation, any ($d-p$)-form) without introducing any gauge fixing conditions or second-class constraints whatsoever as has heretofore always been the case in the literature. Section~\ref{sec:Quantization} addresses the Dirac quantisation of these systems, with the identification of the spectrum of physical states through a likewise factorisation extended to the space of quantum states. Finally, Sect.~\ref{sec:Projection} discusses how the factorisation leads to a most transparent understanding of the projection from TMGT onto TQFT in whatever spacetime dimension in the limit of an infinite topological mass.

\section{Gauge Invariant Factorisation of the Classical Theory}
\label{sec:Classical}

\subsection{Hamiltonian formulation}

Because of the built-in gauge invariances of these systems, the analysis of the constraints \cite{Govaerts:1991,Henneaux:1992} of topologically massive gauge theories is required in order to identify their Hamiltonian formulation. Given the total action (\ref{def:TMGT_Action}) written out in component form the associated Lagrangian density reads,
\begin{eqnarray} \label{def:BF_Lagrangian}
\mathcal{L}_\textrm{TMGT} & = & \frac{\sqrt{h}}{2\, e^2} \, \frac{(-1)^p}{(p+1)!} \, F_{\mu_1\cdots\mu_{p+1}} \,  					F^{\mu_1\cdots\mu_{p+1}}
                            + \frac{\sqrt{h}}{2\, g^2} \, \frac{(-1)^{d-p}}{(d-p+1)!} \, H_{\nu_1\cdots\nu_{d-p+1}} \, H^{\nu_1\cdots\nu_{d-p+1}} \nonumber\\
			  & + & \kappa \, \frac{(1-\xi)}{(1+p)!\, (d-p)!} \, \epsilon^{\mu_1\cdots\mu_{p+1}\nu_1\cdots\nu_{d-p}}  \, F_{\mu_1\cdots\mu_{p+1}} \, B_{\nu_1\cdots\nu_{d-p}} \nonumber\\
			  & - & \kappa \, \frac{\xi\, (-1)^p}{p!\, (d-p+1)!} \, \epsilon^{\mu_1\cdots\mu_p\nu_1\cdots\nu_{d-p+1}}  \, A_{\mu_1\cdots\mu_p} \, H_{\nu_1\cdots\nu_{d-p+1}} ,
\end{eqnarray}
where Greek indices, $\mu,\nu=0,1,\ldots,d$, denote the coordinate indices of the spacetime manifold $\mathcal{M}$ while $h$ is the absolute value of determinant of the metric. According to our conventions, the components of the field strength tensors are given as 
\begin{equation} \label{def:TMGT_Strength}
F_{\mu_1\cdots\mu_{p+1}} = \frac{1}{p!}\, \partial_{\left[ \mu_1 \right.}A_{\left. \mu_2\cdots\mu_{p+1} \right]},
\qquad H_{\nu_1\cdots\nu_{d-p+1}} = \frac{1}{(d-p)!}\, \partial_{\left[ \nu_1 \right.}B_{\left. \nu_2\cdots\nu_{d-p+1} \right]}  ,
\end{equation}
where square brackets on indices denote total antisymmetrisation.
The above expression with the single parameter $\kappa$ multiplying each of the topological $B \wedge F$ and $A \wedge H$ terms while $\xi$ parametrises a possible surface term does not entail any loss of generality. Had two independent parameters $\kappa$ and $\lambda$ multiplying each of the topological terms been introduced, only their sum, $(\kappa+\lambda)$, would have been physically relevant, the other combination corresponding in fact to a pure surface term.

In order to proceed with the Hamiltonian analysis, the spacetime manifold $\mathcal{M}$ is now taken to have the topology of $\mathcal{M}=\mathbb{R} \times \Sigma$ where $\Sigma$ is a compact orientable $d$-dimensional Riemannian space manifold without boundary. Adopting then synchronous coordinates on $\mathcal{M}$, the spacetime metric takes the form $\ud s^2 = \ud t^2 - \tilde{h}_{ij}\, \ud x^i\, \ud x^j $, $\tilde{h}_{ij}(\vec{x}\,)$ being the Riemannian metric on $\Sigma$. Here Latin indices, $i = 1,\ldots,d$, label the space directions in $\Sigma$. The configuration space variable $A(t,\vec{x}\,)$ may then be separated into its temporal component $\ud t \wedge A_0(t,\vec{x}\,)$ with $A_0(t,\vec{x}\,)$ being a $(p-1)$-form on $\Sigma$, and its remaining components $\tilde{A}(t,\vec{x}\,)$ restricted to $\Omega^p(\Sigma)$,
\begin{eqnarray} \label{def:Form_Lorentz}
A_0(t,\vec{x}\,) & = & \frac{1}{(p-1)!} \, A_{0 i_1 \cdots i_{p-1}}(t,\vec{x}\,) \, \ud x^{i_1}\wedge\ldots\wedge \ud x^{i_{p-1}}, \nonumber\\
\tilde{A}(t,\vec{x}\,) & = & \frac{1}{p!} \, A_{i_1 \cdots i_{p}}(t,\vec{x}\,) \, \ud x^{i_1}\wedge\ldots\wedge \ud x^{i_p} .
\end{eqnarray}
A similar decomposition applies to the $(d-p)$-form $B(t,\vec{x}\,)$.

The actual phase space variables are then the spatial components $\tilde{A}$ and $\tilde{B}$ along with their conjugate momenta $\tilde{P}$ and $\tilde{Q}$ defined to be the following differential forms on $\Sigma$,
\begin{eqnarray}\label{def:TMGT_Momenta_Lorentz}
   \tilde{P} & = & \frac{1}{p!}\frac{1}{\sqrt{h}}\, \tilde{h}_{i_1 j_1} \ldots \tilde{h}_{i_p j_p} \, P^{i_1 \cdots i_p} \, \ud x^{j_1} \wedge\ldots\wedge \ud x^{j_p}, \nonumber \\
   \tilde{Q} & = & \frac{1}{(d-p)!}\frac{1}{\sqrt{h}}\, \tilde{h}_{i_1 j_1} \ldots \tilde{h}_{i_{d-p} j_{d-p}}\, Q^{i_1 \cdots i_{d-p}} \, \ud x^{j_1} \wedge\ldots\wedge \ud x^{j_{d-p}} ,
\end{eqnarray}
of which the pseudo-tensorial space components are $P^{i_1 \cdots i_p}$ and $Q^{i_1 \cdots i_{d-p}}$.
Expressed in terms of the configuration space variables, these latter quantities are given as
\begin{eqnarray} \label{def:TMGT_Momenta}
P^{i_1\cdots i_p} & = & \frac{\sqrt{h}}{e^2} \, F_{0 j_1\cdots j_p} \, \tilde{h}^{i_1 j_1} \ldots \tilde{h}^{i_p j_p} + \kappa \, \frac{(1-\xi)}{(d-p)!} \, \epsilon^{i_1 \cdots i_p j_1 \cdots j_{d-p}} \, B_{j_1 \cdots j_{d-p}}, \nonumber \\
Q^{i_1\cdots i_{d-p}} & = & \frac{\sqrt{h}}{g^2} \, H_{0 j_1\cdots j_{d-p}} \, \tilde{h}^{i_1 j_1} \ldots \tilde{h}^{i_{d-p} j_{d-p}} - \kappa \, \frac{\xi}{p!} \, (-1)^{p (d-p)} \, \epsilon^{i_1 \cdots i_{d-p} j_1 \cdots j_p} A_{j_1 \cdots j_p} ,
\end{eqnarray}
while the symplectic structure of Poisson brackets is characterised by the canonical brackets
\begin{eqnarray}
   \left\lbrace A_{i_1\cdots i_p}(t,\vec{x}\,) , P^{j_1\cdots j_p}(t,\vec{y}\,) \right\rbrace &=&  \delta_{\left[ i_1\right. }^{j_1} \ldots \delta_{\left. i_p\right]}^{j_p} \, \delta^{(d)}(\vec{x} - \vec{y}\,), \nonumber \\
   \left\lbrace B_{i_1\cdots i_{d-p}}(t,\vec{x}\,) , Q^{j_1\cdots j_{d-p}}(t,\vec{y}\,) \right\rbrace &=&  \delta_{\left[ i_1\right. }^{j_1} \ldots \delta_{\left. i_{d-p}\right] }^{j_{d-p}} \, \delta^{(d)}(\vec{x} - \vec{y}\,)  .
\end{eqnarray}
{\it A priori\/}, phase space also includes the canonically conjugate variables $A_0$ and $P^0$, and $B_0$ and $Q^0$.

The Legendre transform of the Lagrangian (\ref{def:BF_Lagrangian}) leads to the total gauge invariant Hamiltonian,
\begin{eqnarray} \label{def:TMGT_Hamiltonian}
   H & = & \frac{e^2}{2}\, \left(\ast\tilde{P}-\kappa\, (1-\xi)\, \tilde{B}\right)^2 + \frac{1}{2\, e^2}\, \left(\ud \tilde{A}\right)^2 + \left( u, P^0 \right) \nonumber \\
	      & + & \frac{g^2}{2}\, \left(\ast \tilde{Q}+\kappa\, \xi\, (-1)^{p(d-p)}\tilde{A}\right)^2 + \frac{1}{2\, g^2}\, \left(\ud \tilde{B}\right)^2 \nonumber + \left( v, Q^0 \right) + (\textrm{surface term}) \nonumber\\
& + & \int_{\Sigma} (-1)^p\, (u' + A_0)\wedge\ud\left(\ast\tilde{P}+\kappa\, \xi\, \tilde{B}\right) \nonumber \\
& + & \int_{\Sigma} (-1)^{d-p}  (v' + B_0)\wedge\ud\left(\ast\tilde{Q}-\kappa\, (1-\xi)\, (-1)^{p(d-p)}\, \tilde{A}\right) .
\end{eqnarray}
In this expression as well as throughout hereafter, the Hodge $\ast$ operation is now considered only on the space manifold $\Sigma$ endowed with
the Riemannian metric $\tilde{h}_{ij}$. In (\ref{def:TMGT_Hamiltonian}) the inner product on $\Omega^k(\Sigma)\times\Omega^k(\Sigma)$ is
constructed as
\begin{equation} \label{def:inner_prod}
    (\omega_k)^2 = \left(\omega_k,\omega_k\right) \quad \textrm{with}\qquad
\left(\omega_k,\eta_k\right) = \int_{\Sigma} \omega_k\wedge\ast\eta_k  .
\end{equation}
The quantities $u'$ and $v'$ are Lagrange multipliers for the two first-class constraints associated to abelian gauge symmetries while $u$ and $v$ are those for the first-class constraints $P^0 = 0$ and $Q^0=0$ arising because the fields $A_0$ and $B_0$ are auxiliary degrees of freedom of which the time derivatives do not contribute to the action. Upon reduction to the basic layer of the Hamiltonian nested structure \cite{Govaerts:1991}, $P^0$ and $Q^0$ decouple from the system whereas $A_0$ and $B_0$ play the role of Lagrange multipliers enforcing the two Gauss laws. These constraints generate those gauge transformations in (\ref{def:BF_gauge}) which are continuously connected to the identity transformation, namely the so-called small gauge symmetries, one generated by the fields $\tilde{P}$ and $\tilde{B}$ and the other by $\tilde{A}$ and $\tilde{Q}$, respectively. Note that given Hodge duality, the phase space variables are associated to isomorphic spaces, $\Omega^p(\Sigma) \equiv \Omega^{d-p}(\Sigma)$. Hence at any given spacetime point, phase space has dimension $4 C^p_d$.

\subsection{The Physical-Topological (PT) factorisation}

The above results are well-known. However the fields used to construct the theory do not necessarily create physical states since these are not gauge invariant variables. Therefore, let us now introduce the new Physical-Topological factorisation of the classical theory, by also requiring that these field redefinitions are canonical and preserve canonical commutation relations. First consider the quantities
\begin{equation} \label{def:Tsec_variables}
   \mathcal{A} = -\frac{1}{\kappa} \, (-1)^{p(d-p)} \, \ast\tilde{Q}+(1-\xi)\, \tilde{A} , \qquad
 \mathcal{B} = \frac{1}{\kappa}\ast\tilde{P}+\xi\, \tilde{B}  ,
\end{equation}
defined on the dual sets $\Omega^p(\Sigma)$ and $\Omega^{d-p}(\Sigma)$. This choice is made in such a way that the two Gauss laws are expressed in term of these variables only, as is the case for a topological $B\wedge F$ theory,
\begin{equation} \label{def:TMGT_Gauss}
    \kappa\, (-1)^{p(d-p)}\, \ud\mathcal{A} = 0 , \qquad  (-1)^p\, \kappa\, \ud\mathcal{B} = 0  .
\end{equation}
As a matter of fact, these variables are canonically conjugate,
\begin{equation}\label{eq:Tsec_Bracket}
   \left\lbrace \mathcal{A}_{i_1\cdots i_p}(t,\vec{x}\,) , \mathcal{B}_{j_1\cdots j_{d-p}}(t,\vec{y}\,) \right\rbrace = \frac{1}{\kappa}\, \epsilon_{i_1\cdots i_p j_1\cdots j_{d-p}} \, \delta^{(d)}(\vec{x}-\vec{y}\,)  .
\end{equation}
The two finite gauge transformations in (\ref{def:BF_gauge}) act on these new variables according to the relations,
\begin{equation} \label{eq:Tsec_gauge}
\mathcal{A}' = \mathcal{A} + \alpha , \qquad \mathcal{B}' = \mathcal{B} + \beta .
\end{equation}
At a given spacetime point, these canonically conjugate variables carry $2 C^p_d$ degrees of freedom. The remaining $2 C^p_d$ degrees of freedom are associated to the following pair of gauge invariant variables,
\begin{equation} \label{def:Psec_variables}
   G = \tilde{Q}+\kappa\, \xi\, \ast\tilde{A} , \qquad
   E = \tilde{P}-\kappa\, (1-\xi)\, (-1)^{p(d-p)}\, \ast\tilde{B}  .
\end{equation}
Their pseudo-tensor Lorentz components are defined as in (\ref{def:TMGT_Momenta_Lorentz}) while they possess the following non vanishing canonical Poisson brackets,
\begin{equation} \label{eq:Psec_Bracket}
   \left\lbrace E^{i_1\cdots i_p}(t,\vec{x}\,) , G^{j_1\cdots j_{d-p}}(t,\vec{y}\,) \right\rbrace = -\kappa \, \epsilon^{i_1\cdots i_p j_1\cdots j_{d-p}} \, \delta^{(d)}(\vec{x}-\vec{y}\,)  .
\end{equation}
When considered in combination with the equations of motion, these variables correspond to the non commutative electric fields associated, respectively, to the field strength tensors of $A$ and $B$, see (\ref{def:TMGT_Strength}). Consequently, we have achieved a coherent reparametrisation of phase space which, in fact, factorises the system into two orthogonal sectors, namely sectors of which mutual Poisson brackets vanish identically,
\begin{eqnarray}
\left\lbrace \mathcal{A}_{i_1\cdots i_p}(t,\vec{x}\,) , E^{j_1\cdots j_p}(t,\vec{y}\,) \right\rbrace = 0, &\quad& \left\lbrace \mathcal{A}_{i_1\cdots i_p}(t,\vec{x}\,) , G^{j_1\cdots j_{d-p}}(t,\vec{y}\,) \right\rbrace = 0, \nonumber\\
\left\lbrace \mathcal{B}_{i_1\cdots i_{d-p}}(t,\vec{x}\,) , E^{j_1\cdots j_p}(t,\vec{y}\,) \right\rbrace = 0, &\quad& \left\lbrace \mathcal{B}_{i_1\cdots i_{d-p}}(t,\vec{x}\,) , G^{j_1\cdots j_{d-p}}(t,\vec{y}\,) \right\rbrace = 0 . \nonumber
\end{eqnarray}
Finally in order to obtain the basic nested Hamiltonian formulation \cite{Govaerts:1991} within the factorised parametrisation, the Lagrange multipliers in (\ref{def:TMGT_Hamiltonian}) may be redefined in a convenient way as
\begin{eqnarray}
   u = \dot{A}_0 , &\quad& \mathcal{A}_0 = A_0 + u' + \frac{(-1)^{(p-1)(d-p)}}{2\, g^2\, \kappa^2} \ast \ud \left( \kappa\, \mathcal{B} - 2\ast E \right) , \nonumber\\
   v = \dot{B}_0 , &\quad& \mathcal{B}_0 = B_0 + v' + \frac{(-1)^p}{2\, e^2\, \kappa^2} \ast \ud \left( \kappa\, \mathcal{A} + (1)^{p(d-p)}\, 2\, \ast G \right) , \nonumber
\end{eqnarray}
where a dot stands for differentiation with respect to the time coordinate, $t \in \mathbb{R}$. Consequently the basic total first-class Hamiltonian of the system reads,
\begin{eqnarray} \label{def:Psec_Ham}
   H[E,G,\mathcal{A},\mathcal{B}] & = & \frac{e^2}{2}\, \left(E\right)^2 + \frac{1}{2\, \kappa^2\, g^2}\, \left(\ud^\dagger E\right)^2 + \frac{g^2}{2}\, \left(G\right)^2 + \frac{1}{2\, e^2 \, \kappa^2}\, \left(\ud^\dagger G \right)^2 \nonumber \\
   & + & \kappa\, \int_{\Sigma} (-1)^p\, \mathcal{A}_0\wedge\ud\mathcal{B} - (-1)^{(p+1)(d-p)}\, \mathcal{B}_0\wedge\ud\mathcal{A} ,
\end{eqnarray}
where $\ud^\dagger = *\ud*$ is the coderivative operator. Obviously, $\mathcal{A}_0$ and $\mathcal{B}_0$ are Lagrange multipliers enforcing the first-class constraints which generate the small gauge transformations in (\ref{eq:Tsec_gauge}),
\begin{equation} \label{def:Gauss_constraints}
   \mathcal{G}^{(1)} = \ud\mathcal{A} \qquad \mathcal{G}^{(2)} = \ud\mathcal{B}  .
\end{equation}
When restricted to the physical subspace for which these constraints are satisfied, the above gauge invariant Hamiltonian reduces to a functional depending only on the dynamical physical sector, given by the expression in the first line of (\ref{def:Psec_Ham}).

These redefinitions of the phase space variables have indeed achieved the announced factorisation. A first sector is comprised of the variables constructed in (\ref{def:Tsec_variables}), which decouple from the physical Hamiltonian and are therefore non propagating degrees of freedom. Furthermore, the canonically conjugate variables $\mathcal{A}$ and $\mathcal{B}$ actually share the same Poisson brackets, Gauss law constraints and gauge transformations as the phase space description of a pure $B\wedge F$ topological field theory constructed only from the topological terms in the action (\ref{def:BF_Lagrangian}). Hence this ``topological field theory (TFT) sector" accounts for the $B\wedge F$ theory embedded into the topologically massive gauge theory.

Physical and non physical degrees of freedom are mixed in the original phase space. Our redefinition of fields deals with the original degrees of freedom in such a way that within the Hamiltonian formalism, non propagating (and gauge variant) degrees of freedom are decoupled from the dynamical sector. This latter sector describes only physical degrees of freedom, namely the gauge invariant canonically conjugate electric fields, which diagonalise the physical Hamiltonian (\ref{def:Psec_Ham}) in such a way that they acquire a mass through a mixing of the original variables (\ref{def:Psec_variables}). However the Poisson bracket structure remains unaffected since these field redefinitions define merely a canonical transformation. On account of Hodge duality between $\Omega^p(\Sigma)$ and $\Omega^{d-p}(\Sigma)$, one readily identifies in the dynamical sector the Hamiltonian of a massive $p$-form field of mass $m=\hbar\mu$,
\begin{displaymath}
H[C,E,\mathcal{A},\mathcal{B}] = \frac{\mu^2}{2}\, \left(C\right)^2 + \frac{1}{2}\, \left(\ud C \right)^2 + \frac{1}{2}\, \left(E\right)^2 + \frac{1}{2\, \mu^2}\, \left(\ud E\right)^2 + H_{\textrm{TFT}}[\mathcal{A},\mathcal{B}] .
\end{displaymath}
In comparison with (\ref{def:Psec_Ham}) the following identifications have been applied,
\begin{displaymath}
   \mu=|\kappa\, e\, g|\, \qquad E \to \frac{E}{e} \qquad \ast G = e\, \kappa\, (-1)^{p(d-p)}\, C ,
\end{displaymath}
where $C$ is a $p$-form field of which the Lorentz components are covariant in the manner of (\ref{def:Form_Lorentz}). Physical phase space is then endowed with the elementary Poisson brackets
\begin{displaymath}
\left\lbrace C_{i_1\cdots i_p}(t,\vec{x}\,) , E^{j_1\cdots j_p}(t,\vec{y}\,) \right\rbrace = \delta_{\left[ i_1\right. }^{j_1} \ldots \delta_{\left. i_p\right]}^{j_p} \, \delta^{(d)}(\vec{x} - \vec{y}\,) \, .
\end{displaymath}
Alternatively one may also obtain the Hamiltonian of a massive $(d-p)$-form field of mass $m=\hbar\mu$,
\begin{displaymath}
H[C,G,\mathcal{A},\mathcal{B}] = \frac{\mu^2}{2}\, \left(C\right)^2 + \frac{1}{2}\, \left(\ud C \right)^2 + \frac{1}{2}\, \left(G\right)^2 + \frac{1}{2\, \mu^2}\, \left(\ud G\right)^2 + H_{\textrm{TFT}}[\mathcal{A},\mathcal{B}] ,
\end{displaymath}
in which, in comparison with (\ref{def:Psec_Ham}), the following identifications have been applied,
\begin{displaymath}
   G \to \frac{G}{g} \qquad \ast E = -g\, \kappa\, C .
\end{displaymath}
In this case, $C$ is a ($d-p$)-form field with covariant Lorentz components as in (\ref{def:Form_Lorentz}). The elementary Poisson brackets for these physical phase space variables are
\begin{displaymath}
\left\lbrace C_{i_1\cdots i_{d-p}}(t,\vec{x}\,) , G^{j_1\cdots j_{d-p}}(t,\vec{y}\,) \right\rbrace = \delta_{\left[ i_1\right. }^{j_1} \ldots \delta_{\left. i_{d-p}\right] }^{j_{d-p}} \, \delta^{(d)}(\vec{x} - \vec{y}\,)  .
\end{displaymath}
To conclude this discussion of the factorised Hamiltonian formulation of these TMGT, let us emphasize once more that no gauge fixing procedure whatsoever was applied, in contradistinction to all discussions available until now in the literature leading to an identification of the physical content of these theories. Through the present approach, the TFT content of TMGT is made manifest in a most transparent and simple manner, with in addition a decoupling of the actual physical and dynamical sector of the system from its purely topological one, the latter carrying only topological information characteristic of the underlying spacetime manifold.

\subsection{Hodge decomposition}

The space manifold $\Sigma$ having been assumed to be orientable and compact, let us now consider the consequences of its cohomology group structure, especially in the case when the latter could be non trivial. Throughout the discussion it is implicitly assumed that the $p$- and $(d-p)$-form fields $A$ and $B$ are globally defined differentiable forms in $\Omega^p(\mathcal{M})$ and $\Omega^{d-p}(\mathcal{M})$. When parametrising the theory in terms of the PT factorised variables, the latter assumption of a topological character concerns only the TFT sector. The variables of the dynamical sector are already globally defined whatever the topological properties of the original variables. By virtue of the Hodge theorem \cite{Nakahara:1990}, the phase space variables of the TFT sector, thus globally defined on $\Sigma$ itself endowed with the Riemannian metric $\tilde{h}_{ij}$, may uniquely be decomposed for each time slice into the sum of an exact, a co-exact and a harmonic form with respect to the inner product specified in (\ref{def:inner_prod}),
\begin{equation}\label{def:Hodge_Th}
   \mathcal{A} = A^e + A^c + A_h, \qquad \mathcal{B} = B^e + B^c + B_h .
\end{equation}
A likewise decomposition applies to the dynamical sector. 

Such a decomposition amounts to a split of the fields into a longitudinal part (subscript $L$), a transverse part (subscript $T$) and a ``global" part. The transverse and longitudinal parts are associated to idempotent orthogonal projection operators,
\begin{eqnarray} \label{Hodge_Projection}
   \Pi^T_{(p)} = \frac{1}{\triangle^\bot_{(p)}}\ud^\dagger_{(p+1)} d_{(p)}, &\qquad& 
   \Pi^L_{(p)} = \frac{1}{\triangle^ \bot_{(p)}}\ud_{(p-1)} \ud^{\dagger}_{(p)}, \nonumber \\ 
   \Pi^{T}_{(p)} : \Omega^p(\Sigma) \to (Z^\dagger_\bot)^p(\Sigma), &\qquad& 
   \Pi^{L}_{(p)} : \Omega^p(\Sigma) \to Z_\bot^p(\Sigma) , 
\end{eqnarray}
where $\triangle^\bot_{(p)}$ is the Laplacian operator acting on the space $\Omega^p_\bot(\Sigma)$ of $p$-forms from which the kernel ${\rm ker}\,\triangle_{(p)}$ of the Laplacian $\triangle_{(p)}$ has been subtracted, while $(Z^\dagger_\bot)^p$ (resp. $Z_\bot^p$) is the space of co-closed (resp. closed) $p$-forms non cohomologous to zero. One therefore has the following properties,
\begin{displaymath}
   (-1)^{p(d-p)}\, \Pi^T_{(p)} = \ast \Pi^L_{(d-p)} \ast , \qquad
   \Pi^T_{(p)} + \Pi^L_{(p)} = Id^\bot_{(p)} ,
\end{displaymath}
where $\ast$ is the Hodge star operator on $\Sigma$ and $Id^\bot_{(p)}$ the identity operator on the subspace
$\Omega^p_\bot(\Sigma)$.

In order that the longitudinal and transverse components possess the same physical dimensions as the original fields, the Hodge decomposition of fields may be expressed in terms of a convenient normalisation,
\begin{equation}\label{def:Hodge_normal}
   \sqrt{\triangle^\bot}\mathcal{A}=\ud A_L + \ud^\dagger A_T, \qquad
   \sqrt{\triangle^\bot}\mathcal{B}=\ud B_L + \ud^\dagger B_T .
\end{equation}
Let us then define a new set of variables in the TFT sector, using the projection operators (\ref{Hodge_Projection}),
\begin{eqnarray}\label{def:Hodge_variables}
   \varphi = \Pi^T_{(p-1)}\, A_L , &\quad& \ast Q_{\vartheta} = \Pi^L_{(p+1)} A_T , \nonumber\\
   \vartheta = \Pi^T_{(d-p-1)} B_L , &\quad& \ast P_{\varphi} = \Pi^L_{(d-p+1)} B_T \, , 
\end{eqnarray}
where the components of $\ast P_{\varphi}$ and $\ast Q_{\vartheta}$ are pseudo-tensors defined in a manner analogous to the conjugate momenta in (\ref{def:TMGT_Momenta_Lorentz}). In terms of these new variables the non vanishing Poisson brackets are
\begin{eqnarray}
\left\lbrace \varphi_{i_1 \cdots i_{p-1}}(t,\vec{x}\,), P_\varphi^{j_1 \cdots j_{p-1}}(t,\vec{y}\,) \right\rbrace & = & \frac{1}{\kappa}\, \left(\Pi^T\right)_{i_1 \cdots i_{p-1}}^{j_1 \cdots j_{p-1}}\, \delta^{(d)}(\vec{x}-\vec{y}\,), \nonumber \\ 
\left\lbrace \vartheta_{i_1 \cdots i_{d-p-1}}(t,\vec{x}\,) , Q_\vartheta^{j_1 \cdots j_{d-p-1}}(t,\vec{y}\,) \right\rbrace & = & -\frac{1}{\kappa}\, \left(\Pi^T\right)_{i_1 \cdots i_{d-p-1}}^{j_1 \cdots j_{d-p-1}}\, \delta^{(d)}(\vec{x}-\vec{y}\,)  . \nonumber
\end{eqnarray}
In conclusion, in the TFT sector, rather than working in terms of the phase space variables $\mathcal{A}$ and $\mathcal{B}$
one may parametrise these degrees of freedom in terms of the ``longitudinal" fields $\varphi$ and $\vartheta$ as well as their conjugate momenta, namely the ``transverse" fields $P_\varphi$ and $Q_\vartheta$, to which the harmonic components $A_h$ and $B_h$ of $\mathcal{A}$ and $\mathcal{B}$ must still be adjoined. The same procedure may be applied to the variables of the dynamical sector. The Hamiltonian (\ref{def:Psec_Ham}) then decomposes into a transverse, a longitudinal and a harmonic contribution from these latter variables only.

A natural consequence of the Hodge decomposition is the isomorphism between the $p^\textrm{th}$ de Rham cohomology group, $H^p(\Sigma,\mathbb{R})$, and the space of harmonic $p$-forms, $\ker \triangle_{(p)}$. This means that each equivalence class of $H^p(\Sigma,\mathbb{R})$ has an unique harmonic $p$-form representative identified through the inner product (\ref{def:inner_prod}). It is possible to choose a basis for $\ker \triangle_{(p)}$ in such a way that the harmonic component of any $p$-form is expressed in a topological invariant way. This may be achieved by defining a topological invariant isomorphism between the components of an equivalence class of the $p^\textrm{th}$ (singular) homology group $H_p(\Sigma,\mathbb{R})$ and the components of a form in $\ker \triangle_{(p)}$ (the $p$-homology group is the set of equivalence classes of $p$-cycles differing by a $p$-boundary). Thus, instead of constructing the basis from the Hodge decomposition inner product (\ref{def:inner_prod}), one uses the bilinear, non degenerate and topological invariant inner product $\Lambda$ defined by
\begin{equation}\label{def:dual_product}
   \Lambda :\quad H_p(\Sigma)\times H^p(\Sigma) \to \mathbb{R} :
\quad \Lambda\left(\left[ \Gamma \right],\left[ \omega \right]\right) = \int_{\Gamma} \omega  ,
\end{equation}
making explicit the Poincar\'e duality between homology and cohomology groups \cite{Nakahara:1990}. Given the Hodge theorem, this inner product naturally induces a topological invariant inner product between the equivalent classes of $H_p(\Sigma)$ and the elements of $\ker \triangle_{(p)}$. Therefore, if one introduces generators of the free abelian part of the $p^\textrm{th}$ singular homology group of rank $N_p$, $\left\lbrace \Sigma_{(p)}^\gamma \right\rbrace_{\gamma=1}^{N_p}$, a convenient dual basis $\left\lbrace X^\gamma \right\rbrace $ of $\ker \triangle_{(p)}$ may be chosen such that
\begin{displaymath}
   \Lambda \left(\left[\Sigma_{(p)}^\alpha \right], X^\beta \right) = \delta^{\alpha\beta}  .
\end{displaymath}
Using the duality (\ref{def:dual_product}), the harmonic component $A_h$ of the $p$-form variable $\mathcal{A}$ is thus decomposed according to
\begin{displaymath}
A_h = \sum_{\gamma=1}^{N_p} \Lambda\left(\left[\Sigma_{(p)}^\gamma\right],A_h\right)\, X^\gamma \, .
\end{displaymath}
These components of $A_h$ in the basis $\left\lbrace X^\gamma \right\rbrace $ are topological invariants because they express the periods of $\mathcal{A}$ over the cycle generators of $H_p(\Sigma)$. This is thus nothing other than the classical Wilson loop argument over these generators,
\begin{equation}\label{def:Harm_var(p)}
   a^\gamma = \oint\limits_{\Sigma_{(p)}^\gamma} A_h \, .
\end{equation}
In other words, the variables $a^\gamma(t)$ specify the complete set of remaining ``global" degrees of freedom in the TFT sector for the field $\mathcal{A}$,
\begin{displaymath}
A_h(t,\vec{x}\,)=\sum_{\gamma=1}^{N_p}\,a^\gamma(t)\,X^\gamma(\vec{x}\,).
\end{displaymath}
In a likewise manner, the harmonic component of the ($d-p$)-form variable $\mathcal{B}$ may be decomposed according to
\begin{displaymath}
B_h = \sum_{\gamma=1}^{N_p} \Lambda\left(\left[\Sigma_{(d-p)}^\gamma\right],B_h\right) \, Y^\gamma  ,
\end{displaymath}
where $\left\{Y^\gamma\right\}$ is the dual basis of the cycle generators in $H_{d-p}(\Sigma)$, $\left\lbrace \Sigma_{(d-p)}^\gamma \right\rbrace_{\gamma=1}^{N_p}$. Hence, the components of harmonic ($d-p$)-forms are expressed as
\begin{equation} \label{def:Harm_var(d-p)}
  b^\gamma = \oint\limits_{\Sigma_{(d-p)}^\gamma} B_h  ,
\end{equation}
leading to a similar decomposition of the ``global" degrees of freedom for the dual field $\mathcal{B}$,
\begin{displaymath}
B_h(t,\vec{x}\,)=\sum_{\gamma=1}^{N_p}\,b^\gamma(t)\,Y^\gamma(\vec{x}\,).
\end{displaymath}

The Poisson brackets between the above global variables are topological invariants,
\begin{equation}
\left\lbrace a^{\gamma} , b^{\gamma'} \right\rbrace = \frac{1}{\kappa} I^{\gamma\gamma'}  ,
\end{equation}
namely the signed intersection matrix of which each entry is the sum of the signed intersections of the generators of $H_p(\Sigma)$ and $H_{d-p}(\Sigma)$,
\begin{equation}
   I^{\gamma\gamma'} = I\left[\Sigma_{(p)}^\gamma,\Sigma_{(d-p)}^{\gamma'}\right]  .
\end{equation}
Within our approach, we recover the results of \cite{Polychronakos:1989cd,Bergeron:1994ym,Szabo:1999gm} in 2+1 (on the torus), 3+1 and $d$+1 dimensions, respectively.

\subsection{Large and small gauge transformations}

Only the TFT sector is not gauge invariant. Its phase space variables transform exactly like in a pure $B\wedge F$ theory, see (\ref{eq:Tsec_gauge}). Let us recall that in (\ref{eq:Tsec_gauge}), $\alpha$ and $\beta$ are, respectively, closed $p$- and $(d-p)$-forms on $\Sigma$. In the case of a homologically trivial space $\Sigma$ any closed form is also exact. In the case of a homologically non trivial space $\Sigma$, according to the Hodge theorem any closed form $\alpha$ or $\beta$ may uniquely be decomposed (for a given metric structure) into the sum of an exact and a harmonic form. The exact parts of $\alpha$ and $\beta$ define small gauge transformations, generated by the two Gauss law first-class constraints (\ref{def:TMGT_Gauss}).
Given the Hodge decompositions in the TFT sector (\ref{def:Hodge_normal}) and (\ref{def:Hodge_variables}), these constraints, which require that the phase space variables $\mathcal{A}$ and $\mathcal{B}$ of the TFT sector be closed forms, reduce to
\begin{equation}\label{eq:Gauss_Transverse}
    \mathcal{G}^{(1)} = \sqrt{\triangle^\bot}\, \ast Q_\vartheta , \qquad 
    \mathcal{G}^{(2)} = \sqrt{\triangle^\bot}\, \ast P_\varphi .
\end{equation}
Small gauge transformations act only on the exact part of the TFT sector fields by translating them, namely in terms of the longitudinal $(p-1)$- and $(d-p-1)$-form fields defined in (\ref{def:Hodge_variables}),
\begin{displaymath}
   {\varphi}' = \varphi+\alpha_L, \qquad 
{\vartheta}' = \vartheta + \beta_L ,
\end{displaymath}
where $\alpha_L$ and $\beta_L$ are, respectively, the longitudinal $(p-1)$- and $(d-p-1)$-forms defining the exact components of the gauge transformation forms $\alpha$ and $\beta$ through a construction similar to that in (\ref{def:Hodge_variables}).
The harmonic components of $\alpha$ and $\beta$ define the associated large gauge transformations.

The physical classical phase space in the TFT sector is the set of all field configurations $\mathcal{A}$ and $\mathcal{B}$ obeying the first-class constraints setting to zero their transverse degrees of freedom, see (\ref{eq:Gauss_Transverse}), and identified modulo the action of all gauge transformations, whether small or large. Since under small transformations the longitudinal modes $\varphi$ and $\vartheta$ are gauge equivalent to the trivial configuration of vanishing longitudinal fields, like in any pure $B\wedge F$ TFT the physical phase space of the TFT sector, so far for what concerns small gauge symmetries, is thus finite dimensional and isomorphic to the ensemble of harmonic forms defined modulo exact forms,
\begin{equation} \label{def:TFT_phase-space}
   \mathcal{P} = H^p(\Sigma,\mathbb{R}) \oplus H^{d-p}(\Sigma,\mathbb{R}) ,
\end{equation}
where $H^p(\Sigma,\mathbb{R})$ is the $p^\textrm{th}$ de Rham cohomology group. Let us recall that according to Poincar\'e duality, $H^p(\Sigma)$ is isomorphic to $H^{d-p}(\Sigma)$. Hence, whether one considers functionals of harmonic $p$-forms or $(d-p)$-forms is of no consequence. The finite dimension of this group is given by the corresponding Betti number $N_p$ (for example in the case of the torus, $\Sigma=T_d$, $N_p=C^p_d$). The physical phase space of the TFT sector is thus spanned by the global degrees of freedom $a^\gamma(t)$ and $b^\gamma(t)$, which are indeed obviously invariant under all small gauge transformations. However, this phase space is subjected to further restrictions still, stemming from large gauge transformations.

In a manner similar to the above characterisation of the physical phase space in the TFT sector, the modular group is the quotient of the full gauge group by the subgroup of small gauge transformations generated by the first-class constraints, namely essentially the set of large gauge transformations (LGT) defined modulo small gauge transformations. Large gauge transformations cannot be built from a succession of infinitesimal transformations. They correspond to the cohomologically non trivial, namely the harmonic components of $\alpha$ and $\beta$. Rather than requiring strict invariance of the global phase space variables $a^\gamma$ and $b^\gamma$ under large gauge transformations, having in mind compact U(1) abelian gauge symmetries defined in terms of univalued pure imaginary exponential phase factors, the global physical observables to be considered and thus to be required to remain invariant under large gauge transformations are the holonomy or Wilson loop operators of the TFT sector around compact orientable submanifolds $\Sigma_{p}$ and $\Sigma_{d-p}$ in $\Sigma$. The only non trivial Wilson loops are those around homotopically non trivial cycles, namely elements $\left[\Gamma_p\right]$ of $H_p(\Sigma,\mathbb{Z})$ which may be decomposed in the basis $\left\lbrace \Sigma_{(p)}^\gamma \right\rbrace_{\gamma=1}^{N_p} $. Consequently, given the basis of $\ker \triangle_{(p)}$ constructed from (\ref{def:dual_product}) one has the following set of global Wilson loop observables
\begin{eqnarray}
W[\Gamma_{(p)}] & = & \exp\left( i\, \sum_{\gamma=1}^{N_p} \sigma^\gamma\, \oint\limits_{\Sigma_{(p)}^\gamma} \mathcal{A}\right) = \exp\left(i\, \sum_{\gamma=1}^{N_p} \sigma^\gamma\, a^\gamma \right) ,\nonumber\\
W[\Gamma_{(d-p)}] & = & \exp\left( i\, \sum_{\gamma=1}^{N_p} \tilde{\sigma}^\gamma\, \oint\limits_{\Sigma_{(d-p)}^\gamma} \mathcal{B}\right) = \exp\left(i\, \sum_{\gamma=1}^{N_p} \tilde{\sigma}^\gamma\, b^\gamma \right) \nonumber ,
\end{eqnarray}
where $\sigma^\gamma$, $\tilde{\sigma}^\gamma$ are arbitrary integers. Large gauge transformations associated to closed forms $\alpha$ and $\beta$ act on the global variables $a^\gamma$ and $b^\gamma$ according to
\begin{equation}
a'^\gamma=a^\gamma+\alpha^\gamma,\qquad
b'^\gamma=b^\gamma+\beta^\gamma,
\end{equation}
where $\alpha^\gamma$ and $\beta^\gamma$ are given by
\begin{displaymath}
\alpha^\gamma=\oint\limits_{\Sigma^\gamma_{(p)}}\alpha,\qquad
\beta^\gamma=\oint\limits_{\Sigma^\gamma_{(d-p)}}\beta .
\end{displaymath}

Although the Wilson loops are constructed on the free abelian homology group $H_p(\Sigma,\mathbb{Z})$, the cohomology group including the large gauge transformation parameters is dual to the singular homology group $H_p(\Sigma,\mathbb{R})$. Hence, the only allowed large gauge transformations correspond to components of the harmonic content of the forms $\alpha$ and $\beta$ which are discrete and quantised,
\begin{equation}\label{def:LGT_integer}
   \alpha^\gamma = \oint\limits_{\Sigma_{(p)}^\gamma} \alpha = 2\pi \, \ell^\gamma_{(p)} , \qquad 
   \beta^\gamma = \oint\limits_{\Sigma_{(d-p)}^\gamma} \beta = 2\pi \, \ell^\gamma_{(d-p)}  .
\end{equation}
Here $\ell^\gamma_{(p)}$ and $\ell^\gamma_{(d-p)}$ are integers which characterise the winding numbers of the large gauge transformations, namely the periods of these transformations around the homology cycle generators. The requirement of gauge invariance of all Wilson loops hence constrains the parameters of large gauge transformations to belong to the dual of the free abelian homology group. As a consequence, finally the physical classical phase space in the TFT sector is the quotient of the de Rham cohomology group $H^p(\Sigma,\mathbb{R})\oplus H^{d-p}(\Sigma,\mathbb{R})$ by the additive lattice group defined by the transformations, 
\begin{displaymath}
a'^\gamma=a^\gamma+2\pi \, \ell^\gamma_{(p)},\qquad
b'^\gamma=b^\gamma+2\pi \, \ell^\gamma_{(d-p)},
\end{displaymath}
namely a finite dimensional compact space having the topology of a torus of dimension $2N_p$.

\section{Canonical Quantisation and Physical States}
\label{sec:Quantization}

\subsection{Physical Hilbert space factorisation}

The BRST formalism offers a powerful and elegant quantisation procedure for TMGT but requires the introduction of ghosts. In  some respects, this formalism has also been used for the definition and characterisation of topological quantum field theories \cite{Birmingham:1991ty}. In a related manner, the path integral quantisation of these theories also brings to the fore the characterisation of topological invariants through concepts of quantum field theory. For example, the two-point correlation function of $B\wedge F$ (and $A\wedge F$) theories provides a quantum field theoretic realisation of the linking number of two surfaces of dimensions $p$ and $(d-p)$ embedded in $\mathcal{M}$ and its path integral representation through the Ray-Singer analytic torsion of the underlying manifold. Notwithstanding these achievements, this paper will not rely on such methods which necessarily require some gauge fixing procedure. Rather, ordinary Dirac canonical quantisation methods will be implemented to unravel the physical content of TMGT. First, this quantisation procedure is best adapted to a condensed matter interpretation. It also enables to deal with large gauge transformations on homologically non trivial manifolds. Second, the new Physical-Topological (PT) factorisation identified within the Hamiltonian formulation independently of any gauge fixing procedure makes canonical quantisation especially attractive.

Canonical quantisation readily proceeds from the correspondence principle, according to which classical Poisson brackets are mapped onto equal time quantum commutation relations for the classical variables which are promoted to linear self-adjoint operators acting on the Hilbert space of quantum states in the Schr\"odinger picture at the reference time $t=t_0$,
\begin{eqnarray}
\left[\hat{\mathcal{A}}_{i_1\cdots i_p}(t_0,\vec{x}\,) , \hat{\mathcal{B}}_{j_1\cdots j_{d-p}}(t_0,\vec{y}\,) \right] & = & \frac{i\,\hbar}{\kappa}\, \epsilon_{i_1\cdots i_p j_1\cdots j_{d-p}}\, \delta^{(d)}(\vec{x}-\vec{y}\,) , \nonumber\\
\left[ \hat{E}^{i_1\cdots i_p}(t_0,\vec{x}\,) , \hat{G}^{j_1\cdots j_{d-p}}(t_0,\vec{y}\,) \right] &=& -\frac{i\,\hbar}{\kappa}\, \epsilon^{i_1\cdots i_p j_1\cdots j_{d-p}} \, \delta^{(d)}(\vec{x}-\vec{y}\,) . \nonumber
\end{eqnarray}
A possible representation of the associated Hilbert space is in terms of functionals $\Psi[\mathcal{A},E]$ with their canonical hermitean inner product defined in terms of the field degrees of freedom $\mathcal{A}(\vec{x}\,)$ and $E(\vec{x}\,)$.

It should be clear that the PT factorisation identified at the classical level extends to the quantum system. The full Hilbert space of the system factorises into the tensor product of two separate and independent Hilbert spaces, each of which is the representation space of the operator algebra of either the gauge invariant dynamical sector or the TFT sector. As a consequence of the complete decoupling of these two sectors, one of which contributes to the physical Hamiltonian only, the other to the first-class constraint operators only, a basis of the space of quantum states may be constructed in terms of a likewise factorisation of wave functionals. Symbolically one has
\begin{displaymath}
   \Psi[\mathcal{A},E] = \Phi[E] \, \Psi[\mathcal{A}] .
\end{displaymath}
The component $\Phi[E]$ associated to the dynamical sector is manifestly gauge invariant and is the only one which contributes to the energy spectrum. The physical Hilbert subspace associated to the TFT sector consists of those states wave functionals $\Psi^P[\mathcal{A}]$ which are invariant both under small gauge transformations, namely which belong to the kernel of the first-class constraint operators generating these transformations, and under the large gauge transformations\footnote{Otherwise, the physical wave functional carries a projective representation of the group of LGT \cite{Szabo:1999gm}.} characterised in the previous Section in terms of their lattice action on harmonic $p$- and $(d-p)$-forms.

When the space manifold $\Sigma$ is topologically trivial, for instance in the case of the hyperplane, quantisation of TMGT does not offer much interest {\it per se\/} besides the free dynamics of the dynamical sector, since the TFT sector then possesses a single gauge invariant quantum state. However in the presence of external sources, or when the space manifold $\Sigma$ does have non trivial topology, new and interesting features arise. In the latter situation, to be addressed hereafter, the finite though multi-dimensional gauge invariant content of the TFT sector, $\Psi^P[\mathcal{A}]$, does not contribute to the energy spectrum. Hence it induces a degeneracy of the energy eigenstates of the complete system. As demonstrated later, this degeneracy is restricted by gauge invariance under large gauge transformations (LGT). Since the physical wave functional $\Psi^P[\mathcal{A}]$ in the TFT sector coincides with that of a pure topological quantum field theory, one recovers the results of R.~J.~Szabo \cite{Szabo:1999gm} who solved in the Schr\"odinger picture the pure topological $B\wedge F$ theory (as well as in the presence of sources) in any dimension.

\subsection{The topological sector: Gauss' constraints and LGT}

\subsubsection{Hilbert space and holomorphic polarisation}

At the classical level, phase space has been separated into two decoupled sectors: the TFT and the dynamical sectors. According to the Hodge decomposition theorem (\ref{def:Hodge_Th}), each of the corresponding fields may in turn be decomposed into three further subsectors in terms of their longitudinal, transverse and global components. The Gauss law constraints in conjunction with invariance under small gauge transformations reduce the TFT sector to its global variables only, characterised by the vector space $\mathcal{P}$ of the de Rham cohomology group in (\ref{def:TFT_phase-space}), which is to be restricted further into a compact torus by the lattice action of the appropriate discrete large gauge transformations. Likewise in the dynamical sector, the global degrees of freedom of phase space are also purely topological and are again isomorphic to the $2N_p$-dimensional symplectic vector space $\mathcal{P}$ in (\ref{def:TFT_phase-space}). In each case, these spaces are spanned by the global variables defined as in (\ref{def:Harm_var(p)}) and (\ref{def:Harm_var(d-p)}), namely $(a^\gamma,b^\gamma)$ and $(E^\gamma,G^\gamma)$, respectively. It is quite natural to introduce for these even dimensional vector spaces a complex structure parametrised by a $N_p\times N_p$ complex symmetric matrix, $\tau = \Re(\tau) + i\, \rho$, such that $(-\tau)$ takes its values in the Siegel upper half-space. Such a complex structure introduced over the phase space of global degrees of freedom enables the definition of a holomorphic phase space polarisation, hence quantisation of these sectors.

The same decomposition in terms of longitudinal, transverse and global degrees of freedom applies at the quantum level. Through the correspondence principle, these three subsectors of quantum operators obey the Heisenberg algebra, whether for the TFT or dynamical sector. Let us presently restrict to the TFT sector. For what concerns the local operators, one has
\begin{eqnarray} \label{eq:TFT_Commut_1-2}
\left[\hat{\varphi}_{i_1\cdots i_{p-1}}(t_0,\vec{x}\,),\hat{P}_\varphi^{j_1\cdots j_{p-1}}(t_0,\vec{y}\,) \right] 
   & = & \frac{i\, \hbar}{\kappa}\, \left(\Pi^T\right)_{i_1 \cdots i_{p-1}}^{j_1 \cdots j_{p-1}}\, \delta^{(d)}(\vec{x}-\vec{y}\,), \nonumber \\ 
\left[\hat{\vartheta}_{i_1 \cdots i_{d-p-1}}(t_0,\vec{x}\,) , \hat{Q}_\vartheta^{j_1 \cdots j_{d-p-1}}(t_0,\vec{y}\,)\right] 
   & = & -\frac{i\, \hbar}{\kappa}\, \left(\Pi^T\right)_{i_1 \cdots i_{d-p-1}}^{j_1 \cdots j_{d-p-1}}\, \delta^{(d)}(\vec{x}-\vec{y}\,) ,
\end{eqnarray}
while for the global operators,
\begin{displaymath}
\left[ \hat{a}^{\gamma}(t_0) , \hat{b}^{\gamma'}(t_0) \right] = i\, \frac{\hbar}{\kappa}\, I^{\gamma\gamma'} .
\end{displaymath}
Introducing now the holomorphic combinations of the latter operators\footnote{It is implicitly assumed here that the parameter $\kappa$ is strictly positive. If $\kappa$ is negative, the roles of the operators $\hat{a}^\gamma$ and $\hat{b}^\gamma$ are simply exchanged in the discussion hereafter.},
\begin{equation}\label{def:holomorphic_param}
   \hat{c}_\gamma = \sqrt{\frac{\kappa}{2\, \hbar}}\, \sum^{N_p}_{\delta=1} \left( I_{\gamma \delta}\, \hat{a}^\delta + \tau_{\gamma\delta}\, \hat{b}^\delta \right), \qquad
   \hat{c}^\dag_\gamma = \sqrt{\frac{\kappa}{2\, \hbar}}\, \sum^{N_p}_{\delta=1} \left( I_{\gamma \delta}\, \hat{a}^\delta + \overline{\tau}_{\gamma\delta}\, \hat{b}^\delta \right) ,
\end{equation}
where $I_{\gamma\delta}$ is the inverse of the intersection matrix,
\begin{displaymath}
\sum^{N_p}_{\delta=1} I_{\gamma\delta}\, I^{\delta\gamma'} = \delta_\gamma^{\gamma'} ,
\end{displaymath}
one finds the Fock type algebra
\begin{equation}\label{eq:TFT_Commut_Holom}
\left[ \hat{c}_\gamma,\hat{c}^\dag_{\gamma'} \right] = \Im(\tau)_{\gamma{\gamma'}} =\rho_{\gamma{\gamma'}} ,
\end{equation}
all other possible commutators vanishing identically. Note that this result implies that the inner product in this sector of Hilbert space is to be defined in terms of the imaginary part $(\rho^{-1})^{\gamma\gamma'}$, in a manner totally independent from the Riemannian metric structure of the compact space submanifold $\Sigma$. {\it A priori\/}, physical observables in pure topological quantum field theories ought nevertheless to be independent from any extraneous {\it ad hoc\/} structure introduced through the quantisation process such as the present complex structure.

Gauss law constraints and large gauge transformations are to be considered in the wave functional representation of Hilbert space. The latter is spanned by the direct product of basis vectors for the representation spaces of the algebras (\ref{eq:TFT_Commut_1-2}) and (\ref{eq:TFT_Commut_Holom}). These consist of functionals $\Psi[\varphi,\vartheta,c]$ of the infinite dimensional space of field configurations in the TFT sector. Accordingly, the inner product of such states is defined by
\begin{displaymath}
   \left\langle \Psi_1 \right.\left\vert \Psi_2 \right\rangle = \int \left[ \mathcal{D}\varphi \right]\, \left[ \mathcal{D}\vartheta \right]\, \left[\prod_\gamma \ud c_\gamma\right] \, \Psi^\ast_1[\varphi,\vartheta,c]\, \Psi_2[\varphi,\vartheta,c] ,
\end{displaymath}
which requires the specification of a functional integration measure. This measure is taken to be the gaussian measure for fluctuations in the corresponding fields, which is induced by the complex structure $\tau$ in the global sector or else by the Riemannian metric on $\Sigma$ for fluctuations in $\varphi$ and $\vartheta$,
\begin{eqnarray}
\delta \varphi^2 & = & 
   \int_\Sigma \ud\vec{x}\, h^{i_1 k_1}(\vec{x}\,)\ldots h^{i_{p-1}k_{p-1}}(\vec{x}\,)\, \delta \varphi_{i_1\cdots i_{p-1}}(\vec{x}\,)\, \delta \varphi_{k_1\cdots k_{p-1}}(\vec{x}\,), \nonumber\\
\delta \vartheta^2 & = & 
   \int_\Sigma \ud\vec{x}\, h^{j_1 l_1}(\vec{x}\,)\ldots h^{j_{d-p-1}l_{d-p-1}}(\vec{x}\,)\, \delta \vartheta_{j_1\cdots j_{d-p-1}}(\vec{x}\,)\, \delta \vartheta_{l_1\cdots l_{d-p-1}}(\vec{x}\,), \nonumber\\
{\delta c}^2 & = & 
   \sum_{\gamma,{\gamma'}=1}^{N_p} (\rho^{-1})^{\gamma\gamma'} \delta c_\gamma\, \delta c_{\gamma'} \label{def:TFT_metric}  .
\end{eqnarray}
In contradistinction to an ordinary pure topological quantum field theory, such a space metric is readily available within the context of TMGT, being necessary for the specification of the dynamical fields. Independently from the complex structure introduced in the global sector, independence of the physical Hilbert space measure in the $(\varphi,\vartheta)$ sector on the metric on $\Sigma$ will be established hereafter. Consequently the canonical commutation relations (\ref{eq:TFT_Commut_1-2}) and (\ref{eq:TFT_Commut_Holom}) in the TFT sector are represented by the following functional operators acting on the Hilbert space wave functionals,
\begin{eqnarray}
\hat{\varphi}_{i_1 \cdots i_{p-1}}( \vec{x}\,) \equiv \varphi_{i_1 \cdots i_{p-1}}(\vec{x}\,), &\,&
\hat{P}_\varphi^{i_1 \cdots i_{p-1}}(\vec{x}\,) \equiv 
   -\frac{i\, \hbar}{\kappa}\, \left(\Pi^T\right)_{j_1 \cdots j_{p-1}}^{i_1 \cdots i_{p-1}}\, \frac{\delta}{\delta \varphi_{j_1 \cdots j_{p-1}}( \vec{x}\,)} \label{eq:dyn_WF_operator1} ,\\
\hat{\vartheta}_{i_1\cdots i_{d-p-1}}(\vec{x}\,) \equiv \vartheta_{i_1\cdots i_{d-p-1}}(\vec{x}\,), &\,&
\hat{Q}_\vartheta^{i_1 \cdots i_{d-p-1}}(\vec{x}\,) \equiv
   \frac{i\, \hbar}{\kappa}\, \left(\Pi^T\right)_{j_1 \cdots j_{d-p-1}}^{i_1 \cdots i_{d-p-1}}\, \frac{\delta}{\delta \vartheta_{j_1 \cdots j_{d-p-1}}( \vec{x}\,)} \label{eq:dyn_WF_operator2} ,\\
\hat{c}_\gamma \equiv c_\gamma, &\,&
\hat{c}^\dagger_\gamma \equiv 
   - \sum_{\gamma'=1}^{N_p} \rho_{\gamma\gamma'}\, \frac{\partial}{\partial c_{\gamma'}} . \label{eq:dyn_WF_Holom}
\end{eqnarray}

\subsubsection{Gauss law constraints}

The physical Hilbert space is invariant under all gauge transformations. A first restriction arises by requiring the physical quantum states to be invariant under small gauge transformations generated by the first class constraints. This set is the kernel of the Gauss law constraint operators (\ref{def:Gauss_constraints}) which remain defined as in the classical theory since no operator ordering ambiguity is encountered,
\begin{eqnarray}
\hat{\mathcal{G}}^{(1)} \left\vert \Psi^P \right\rangle = 0 & \Rightarrow & 
   \frac{\delta}{\delta \vartheta_{i_1 \cdots i_{d-p-1}}(\vec{x})} \Psi^P[\varphi,\vartheta,c] = 0 , \nonumber\\
\hat{\mathcal{G}}^{(2)} \left\vert \Psi^P \right\rangle = 0 & \Rightarrow & 
   \frac{\delta}{\delta \varphi_{i_1 \cdots i_{p-1}}( \vec{x})} \Psi^P[\varphi,\vartheta,c] = 0 . \nonumber
\end{eqnarray}
Hence physical quantum states necessarily consist of wave functionals which are totally independent of the longitudinal variables ($\varphi,\vartheta$). When restricted to such states and properly renormalised, the inner product integration measure is constructed from the definition (\ref{def:TFT_metric}) of the gaussian metric on the space of fluctuations in the global coordinates,
\begin{displaymath}
   \left\langle \Psi_1 \right.\left\vert \Psi_2 \right\rangle = \int \prod_\gamma \ud c_\gamma\, \left( \det \rho \right)^{-1/2}\,  \Psi^\ast_1(c)\, \Psi_2(c) .
\end{displaymath}
This measure on the physical Hilbert space is thus indeed independent of the Riemannian metric on $\Sigma$, and involves only the {\it ad hoc\/} complex structure $\tau$ introduced towards the quantisation of the global TFT sector.

\subsubsection{LGT and global variables}

The structure of the physical Hilbert space dramatically depends on the way one deals with LGT. Given the holomorphic parametrisation (\ref{def:holomorphic_param}), under the lattice action of LGT of periods $(\ell^\gamma_{(p)},\ell^\gamma_{(d-p)})\equiv(\underline{\ell}_{(p)},\underline{\ell}_{(d-p)})$ as defined in (\ref{def:LGT_integer}) the new global operators should transform as,
\begin{eqnarray}
   c'_\gamma &=& c_\gamma + \sqrt{\frac{2\, \pi^2\, \kappa}{\hbar}} \sum^{N_p}_{\gamma'=1} \left( I_{\gamma \gamma'}\, \ell^{\gamma'}_{(p)} + \tau_{\gamma\gamma'}\, \ell_{(d-p)}^{\gamma'} \right) ,\nonumber \\ 
{c'_\gamma}^\dag &=& c^\dag_\gamma + \sqrt{\frac{2\, \pi^2\, \kappa}{\hbar}} \sum^{N_p}_{\gamma'=1} \left( I_{\gamma \gamma'}\, \ell^{\gamma'}_{(p)} + \overline{\tau}_{\gamma\gamma'}\, \ell_{(d-p)}^{\gamma'} \right) .
\end{eqnarray}
Using the Baker-Campbell-Hausdorff (BCH) formulae for any two operators $\hat{A}$ and $\hat{B}$ commuting with their own commutator,
\begin{equation}
e^{\hat{A}}\, \hat{B}\, e^{-\hat{A}} = \hat{B} + \left[ \hat{A} , \hat{B} \right] , \qquad
e^{\hat{A}+\hat{B}} = e^{-\frac{1}{2} \left[\hat{A} , \hat{B}\right]}\, e^{\hat{A}}\, e^{\hat{B}} ,
\label{def:BCH_2}
\end{equation}
it may be seen that the quantum operator generating the LGT of periods $(\underline{k}_{(p)},\underline{k}_{(d-p)})$ is
\begin{eqnarray} \label{eq:LGT_Operator}
\hat{U}\left( \underline{k}_{(p)},\underline{k}_{(d-p)} \right) & = & C \left( \underline{k}_{(p)},\underline{k}_{(d-p)} \right)\, \prod_{\gamma,{\gamma'},\epsilon}^{N_p} \exp\left\lbrace 2\, \pi\, \sqrt{\frac{\kappa}{2\, \hbar}} \right. \nonumber\\
& & \left. \times (\rho^{-1})^{\gamma{\gamma'}} \left[\left( I_{\gamma\epsilon}\, k^\epsilon_{(p)} + \overline{\tau}_{\gamma\epsilon}\, k^\epsilon_{(d-p)} \right) \hat{c}_{\gamma'} - \left( I_{\gamma\epsilon}\, k^\epsilon_{(p)} + \tau_{\gamma\epsilon}\, k^\epsilon_{(d-p)} \right) \hat{c}^{\dagger}_{\gamma'} \right] \right\rbrace .
\end{eqnarray}
The 1-cocycle $C \left( \underline{k}_{(p)},\underline{k}_{(d-p)} \right)$ will be determined presently. This operator (\ref{eq:LGT_Operator}) defines the action of LGT on the Hilbert space in the global TFT sector,
\begin{eqnarray}\label{eq:LGT_Wavefunction}
\hat{U}\left(\underline{k}_{(p)},\underline{k}_{(d-p)}\right) \Psi \left(c_\gamma \right) &=& \prod_{\gamma,\gamma',\delta}^{N_p} e^{\pi^2\, \frac{\kappa}{\hbar}\, \left[I_{\gamma\delta}\, k^{\delta}_{(p)} + \overline{\tau}_{\gamma\delta}\, k^{\delta}_{(d-p)}\right]\, (\rho^{-1})^{\gamma\gamma'}\, \left[ \sqrt{\frac{2\, \hbar}{\pi^2\, \kappa}}\, c_{\gamma'} + \sum_\epsilon \left\lbrace I_{\gamma'\epsilon}\, k^{\epsilon}_{(p)} + \tau_{\gamma'\epsilon}\, k^{\epsilon}_{(d-p)} \right\rbrace\right]} \nonumber\\
& & C\left(\underline{k}_{(p)},\underline{k}_{(d-p)} \right) \Psi\left(c_\gamma + \pi \sqrt{\frac{2 \kappa}{\hbar}} \sum_{\gamma'=1}^{N_p} \left[ I_{\gamma\gamma'}\, k^{\gamma'}_{(p)} + \tau_{\gamma\gamma'}\, k^{\gamma'}_{(d-p)} \right] \right)  \, ,
\end{eqnarray}
where the BCH formula (\ref{def:BCH_2}) has been used.
However, a U(1)$\times$U(1) 2-cocycle $\omega_2(k;\ell)$ appears in the composition law of this quantum representation,
\begin{eqnarray}
   \hat{U}\left( \underline{k}_{(p)} + \underline{\ell}_{(p)}, \underline{k}_{(d-p)} + \underline{\ell}_{(d-p)} \right) & = & e^{2\, i\, \pi\, \omega_2(k;\ell)}\, \hat{U}\left( \underline{k}_{(p)},\underline{k}_{(d-p)} \right)\, \hat{U}\left( \underline{\ell}_{(p)},\underline{\ell}_{(d-p)} \right) , \nonumber\\
   \omega_2(k;\ell) \equiv \omega_2\left( \underline{k}_{(p)} , \underline{k}_{(d-p)} ; \underline{\ell}_{(p)} , \underline{\ell}_{(d-p)} \right) & = & \frac{\pi\, \kappa}{\hbar}\, \sum_{\gamma,\gamma'=1}^{N_p} I_{\gamma{\gamma'}} \left[\ell^\gamma_{(d-p)}\, k^{\gamma'}_{(p)} - k^\gamma_{(d-p)}\, \ell^{\gamma'}_{(p)} \right] \nonumber .
\end{eqnarray}
The 1-cocycle $C \left( \underline{k}_{(p)},\underline{k}_{(d-p)} \right)$ appearing in (\ref{eq:LGT_Operator}) may be determined by requiring that the abelian group composition law for LGT is recovered. This implies that $\omega_2(k;\ell)$ is a coboundary,
\begin{eqnarray}
   \omega_2(k;\ell) &=& \mathcal{C}_1\left( \underline{k}_{(p)} + \underline{\ell}_{(p)} , \underline{k}_{(d-p)} + \underline{\ell}_{(d-p)} \right) - \mathcal{C}_1\left( \underline{k}_{(p)}, \underline{k}_{(d-p)} \right) - \mathcal{C}_1\left( \underline{\ell}_{(p)} , \underline{\ell}_{(d-p)} \right) \quad (\textrm{mod } \mathbb{Z}), \nonumber\\
   C (k) &\equiv& C \left( \underline{k}_{(p)},\underline{k}_{(d-p)} \right) = e^{2\, i\, \pi\, \mathcal{C}_1\left( \underline{k}_{(p)}, \underline{k}_{(d-p)} \right)}. \nonumber
\end{eqnarray}
A careful analysis, analogous to the one in \cite{Govaerts:1999fg}, finds that the unique solution to this coboundary condition is
\begin{equation} \label{eq:kappa_quant}
   \kappa = \frac{\hbar}{2\pi}\, \mathcal{I}\, k  , \qquad 
   C \left( \underline{k}_{(p)},\underline{k}_{(d-p)} \right) = \prod_{\gamma,\gamma'=1}^{N_p} e^{i\, \pi\, k\, \mathcal{I}\, k^\gamma_{(d-p)}\, I_{\gamma\gamma'}\, k^{\gamma'}_{(p)}} ,
\end{equation}
where $k \in \mathbb{Z}$ and\footnote{Recall that $\kappa$, hence $k$ is assumed to be strictly positive in the present discussion, while the situation for a negative $\kappa$ or $k$ is obtained through the exchange of the sectors $a^\gamma$ and $b^\gamma$.} $\mathcal{I} = \det\left( I^{\gamma\gamma'} \right) \in \mathbb{N}$. It is noteworthy to recall that although $I_{\gamma\gamma'}$ is a rational valued matrix, $\mathcal{I}\, I_{\gamma\gamma'}$ is integer valued. Note also the quantisation condition arising for the coefficient $\kappa$ multiplying the topological terms in the original action of TMGT.

If $k$ is rational, namely if $k=k_1/k_2$ with $k_1 , k_2$ strictly positive natural numbers, invariance of physical states under LGT cannot be achieved. However in this case the LGT group has a finite dimensional projective representation which may be constructed by finding a normal subgroup generated by the LGT operators. As demonstrated in \cite{Szabo:1999gm}, the TFT part of the physical wave functions carries a projective representation of the group of LGT while the above discussion establishes that the dimension of Hilbert space is $\prod_{\delta=1}^{N_p} k_1\, k_2\, \mathcal{I}\, \textrm{Min}(I_{\delta\delta'})$. Any state of a given irreducible representation gives the same matrix element for a physical observable. The characterisation of Hilbert space changes qualitatively for integer or rational values of $k$, but the theory remains well-defined.

If we take $k$ to be an integer, see (\ref{eq:kappa_quant}), wave functions of the physical Hilbert space may be classified in terms of irreducible representations of the group of LGT (\ref{eq:LGT_Wavefunction}),
\begin{eqnarray}
  &&\Psi\left( \eta_1 ; c_\gamma + \sqrt{\pi\, \mathcal{I}\, k}\, \sum_{\gamma'}^{N_p} \left[ I_{\gamma\gamma'}\, k^{\gamma'}_{(p)} + \tau_{\gamma\gamma'}\, k^{\gamma'}_{(d-p)} \right]\right)\nonumber\\ &=&  \prod_{\gamma,\gamma',\delta=1}^{N_p} \exp\left\lbrace -\sqrt{\pi \mathcal{I}\, k} \left[I_{\gamma\delta}\, k^{\delta}_{(p)} + \overline{\tau}_{\gamma\delta}\, k^{\delta}_{(d-p)}\right] (\rho^{-1})^{\gamma\gamma'} \left[ c_{\gamma'} + \frac{\sqrt{\pi \mathcal{I}\, k}}{2} \sum_\epsilon \left\lbrace I_{\gamma'\epsilon}\, k^{\epsilon}_{(p)} + \tau_{\gamma'\epsilon}\, k^{\epsilon}_{(d-p)} \right\rbrace\right]\right\rbrace\nonumber\\
&\times& \prod_{\gamma,\gamma'=1}^{N_p} \exp\left\lbrace 2\, i\, \pi\, \eta_1(\underline{k}_{(p)},\underline{k}_{(d-p)}) - i\, \pi\, k\, \mathcal{I}\, k^\gamma_{(d-p)}\, I_{\gamma\gamma'}\, k^{\gamma'}_{(p)}\right\rbrace\, \Psi\left(\eta_1;c_\gamma\right)  ,
\end{eqnarray}
where the 1-cocycle $\eta_1\left(\underline{k}_{(p)},\underline{k}_{(d-p)}\right)$ characterises the irreducible representation. Since for an abelian group each of its irreducible representations is one-dimensional, physical states corresponding to a given irreducible representation are singlet under LGT.

As is well-known, functions obeying such a double periodicity condition are nothing other than the generalised Riemann theta functions defined in any dimension on the complex $N_p$-torus~\cite{Szabo:1999gm}, with the compact reduced phase space resulting from the requirement of invariance under LGT,
\begin{eqnarray}
\Psi^{r_\delta} \binom{a_\delta}{b_{\delta}} \left( c_\gamma \right) =
\prod_{\gamma,{\gamma'}=1}^{N_p} \left( e^{-\frac{1}{2}\, c_\gamma\, (\rho^{-1})^{\gamma\gamma'}\, c_{\gamma'}} \right)\, \Theta \left[ \begin{array}{c} \sum_{\delta'=1}^{N_p} \frac{I^{\delta'\delta}}{\mathcal{I}\, k} \left( a_{\delta'} + r_{\delta'} \right) \\ b_\delta \end{array} \right] \left( \left. \sqrt{\frac{\mathcal{I}\, k}{\pi}}\, c_\gamma \right| -\mathcal{I}\, k\, \tau \right)  ,
\end{eqnarray}
where $r_\delta \in \left[ 0 , k\, \mathcal{I}\, \textrm{Min}(I_{\delta\delta'}) - 1 \right] \subset \mathbb{N}$.
Each physical subspace, characterised by the 1-cocycle
\begin{displaymath}
\eta^{(a b)}_1(\underline{k}_{(p)},\underline{k}_{(d-p)}) = a_\gamma\, k_{(p)}^\gamma + b_\gamma\, k_{(d-p)}^{\gamma}
\end{displaymath}
where $a_\gamma , b_\gamma \in [0,1[ \subset \mathbb{R}$, is invariant under a particular irreducible representation of LGT. The TFT component of each physical Hilbert space is of dimension $\prod_{\delta=1}^{N_p} k\, \mathcal{I}\, \textrm{Min}(I_{\delta\delta'})$. In general, the choice of physical Hilbert space which is invariant under all LGT is the representation space with $\eta_1(\underline{k}_{(p)},\underline{k}_{(d-p)}) \in \mathbb{Z}$, namely corresponding to $a_\gamma,b_\gamma=0$.

\subsection{The dynamical sector: Hamiltonian diagonalisation}

Based on Hodge's theorem, (\ref{def:Hodge_Th}) and (\ref{def:Hodge_normal}) define the decomposition of the dynamical sector into three decoupled subsectors of canonically conjugate variables: the global harmonic sector and the local $(E_L,P_E)$ and $(G_L,Q_G)$ sectors. In turn the classical Hamiltonian (\ref{def:Psec_Ham}) decomposes into three separate contributions, one for each subsector. When quantising the system in each subsector, the total quantum Hamiltonian follows from the classical one without any operator ordering ambiguity,
\begin{displaymath}
   \hat{H}[\hat{E},\hat{G}] = \hat{H}_h[\hat{E}_h,\hat{G}_h] + \hat{H}_1[\hat{E}_L,\hat{P}_E] + \hat{H}_2[\hat{G}_L,\hat{Q}_E] .
\end{displaymath}
The physical spectrum is thus identified by diagonalising each of these contributions separately.

\subsubsection{Global degrees of freedom}

The choice of normalisation used previously in the harmonic sector relies on the Poincar\'e duality between the basis elements $\left[X^\gamma\right]$ and $\left[Y^\gamma\right]$ of the relevant cohomology groups and their associated homology generators $\Sigma_{(p)}^\gamma$ and $\Sigma_{(d-p)}^\gamma$, respectively, see (\ref{def:dual_product}). This choice is of a purely topological character. However in the dynamical sector, there is a remaining freedom as far as the normalisation of the choice of the harmonic representative of the cohomology group is concerned, depending on the metric structure, and thus fixing the basis elements $X^\gamma$ of $\ker \triangle_{(p)}$ and $Y^\gamma$ of $\ker \triangle_{(d-p)}$. This choice involves the inner product in (\ref{def:inner_prod}) on which the Hodge decomposition relies. Hence one sets
\begin{equation} \label{def:Global_dyn_Metrics}
   \int_\Sigma X_\gamma\wedge\ast X_{\gamma'} = \frac{e}{g}\, \Omega_{\gamma\gamma'} , \qquad 
   \int_\Sigma Y_\gamma\wedge\ast Y_{\gamma'} = \frac{g}{e}\, \tilde{\Omega}_{\gamma\gamma'} ,
\end{equation}
where $\Omega_{\gamma\gamma'}$ and $\tilde{\Omega}_{\gamma\gamma'}$ are $N_p\times N_p$ real symmetric matrices. Given this normalisation, the global part of the metric dependent quantum Hamiltonian operator constructed from (\ref{def:Psec_Ham}) is expressed as
\begin{equation}
   \hat{H}_h[\hat{E}^\gamma,\hat{G}^{\gamma'}] = \frac{1}{2}e\, g\, \sum_{\gamma,\gamma'=1}^{N_p}\left[
 \hat{E}^\gamma\, \hat{E}^{\gamma'}\, \Omega_{\gamma\gamma'} + \hat{G}^\gamma \hat{G}^{\gamma'}\, \tilde{\Omega}_{\gamma\gamma'}\right] ,
\end{equation}
while the non vanishing commutation relations between the global phase space operators read
\begin{equation}
\left[\hat{E}^\gamma , \hat{G}^{\gamma'} \right] = -i\, \hbar\, \kappa\, I^{\gamma\gamma'} .
\end{equation}

As in the TFT sector, see (\ref{def:holomorphic_param}), the following holomorphic polarisation of the global dynamical sector is used,
\begin{displaymath}
d_\gamma = \frac{1}{\sqrt{2\hbar\, \kappa}} \sum^{N_p}_{\alpha=1} \left( I_{\gamma\alpha}\, \hat{E}^\alpha - \upsilon_{\gamma\alpha}\, \hat{G}^\alpha \right) , \qquad
d^\dagger_\gamma = \frac{1}{\sqrt{2\hbar\, \kappa}} \sum^{N_p}_{\alpha=1} \left( I_{\gamma\alpha}\, \hat{E}^\alpha - \overline{\upsilon}_{\gamma\alpha}\, \hat{G}^\alpha \right)  ,
\end{displaymath}
where $\upsilon = \Re(\upsilon) + i\, \sigma$ is the $N_p\times N_p$ complex symmetric matrix characterising the complex structure introduced in the global dynamical phase space sector, of which the imaginary part determines the non vanishing commutation relations of the Fock like algebra
\begin{equation}\label{eq:Global_Dyn_Fock}	
   \left[ d_\gamma,d^\dagger_{\gamma'} \right] = \sigma_{\gamma{\gamma'}}   .
\end{equation}

In order to readily diagonalise the Hamiltonian in the global sector which is of the harmonic oscillator form, it is convenient to make the following choice for the complex structure matrix $v$ as well as for the normalisation quantities specified in (\ref{def:Global_dyn_Metrics}),
\begin{equation} \label{def:Global_dyn_Conventions}
\Re(\upsilon) = 0 , \qquad \sigma_{\gamma\gamma'} = \tilde{\Omega}_{\gamma\gamma'} = \delta_{\gamma\gamma'} , \qquad
\Omega_{\gamma\gamma'} = \sum_{\alpha,\beta=1}^{N_p} I_{\alpha\gamma}\, I_{\beta\gamma'}\, \delta^{\alpha\beta} ,
\end{equation}
where $\delta^{\gamma\gamma'}$ is the $N_p \times N_p$ Kronecker symbol. With these choices, the contribution of the global variables to the Hamiltonian is indeed diagonal,
\begin{displaymath}
H_g =  \frac{1}{2}\, \hbar\, \mu\, N_p + \hbar\, \mu\, \sum_{\gamma,{\gamma'}=1}^{N_p} d^\dagger_\gamma\, d_{\gamma'}\, \delta^{\gamma{\gamma'}},
\qquad \mu=e\,g\,\kappa .
\end{displaymath}
One recognizes the Hamiltonian of a collection of $N_p$ independent harmonic oscillators of angular frequency\footnote{Recall that under the assumptions of the analysis, this combination of parameters is indeed positive.} $\mu = e\, g\, \kappa$, which turns out to be the mass gap of the quantum field theory. The operators $d_\gamma$ and $d^\dagger_\gamma$ are, respectively, annihilation and creation operators obeying the Fock algebra (\ref{eq:Global_Dyn_Fock}) now with $\sigma_{\gamma\gamma'}=\delta_{\gamma\gamma'}$. The energy spectrum in the global dynamical sector of the system is readily identified. The normalised fundamental state is the kernel of all annihilation operators,
\begin{displaymath}
 d_\alpha \left\vert 0 \right\rangle = 0 , \qquad \varepsilon^h_{(0)} = \frac{1}{2}N_p\, \hbar\, \mu ,\qquad
\langle 0|0\rangle =1,
\end{displaymath}
where $\varepsilon^h_{(0)}$ is the vacuum energy. Excited states, $|n_\gamma\rangle$, are obtained through the action of the $N_p$ creation operators $d^\dagger_\gamma$ on the fundamental state. This leads to the energy eigenvalue for any of these states,
\begin{equation} \label{eq:Global_dyn_Energy}
 \left\vert n_\gamma \right\rangle=\prod_{\gamma=1}^{N_p}\frac{1}{\sqrt{n_\gamma !}}\left(d^\dagger_\gamma\right)^{n_\gamma}|0\rangle, \qquad \varepsilon_{(n_\gamma)} = \varepsilon^h_{(0)} + \hbar\, \mu\, \sum_{\gamma=1}^{N_p} n_\gamma  ,
\end{equation}
$\left\lbrace n_\gamma\right\rbrace _{\gamma=1}^{N_p}$ being the eigenvalues of each of the number operators
$d^\dagger_\gamma d_\gamma$, hence positive integers.

\subsubsection{Local degrees of freedom on the torus}

The canonical treatment of the global degrees of freedom in both the TFT and dynamical sectors does not require the explicit specification of the space manifold $\Sigma$ with its topology and Riemannian metric, yet allowing the general discussion of the previous Sections. However, in order to identify the full spectrum of dynamical physical states, the space manifold $\Sigma$ including its geometry has now to be completely specified. The explicit choice to be made for the purpose of the present discussion is that of the $d$-dimensional Euclidean torus, $\Sigma=T_d$, enabling straightforward Fourier mode analysis of the then infinite discrete, thus countable set of degrees of freedom, and diagonalisation of the harmonic oscillator structure of the Hamiltonian. This particular choice of the $d$-torus is motivated by the fact that this manifold is the simplest flat yet homologically non trivial manifold. The notations used are those of \cite{Payen:DEA} where pure quantum electrodynamics is solved on the torus, of which the conventions are extended to any $p$-form. 

Accordingly, the variables $E$ and $G$ of the dynamical sector are periodic around the torus $p$- and $(d-p)$-cycles, respectively. Their Fourier mode expansions read
\begin{displaymath}
   E_\bot^{i_1 \ldots i_p}(\vec{x}\,) = \delta^{i_1j_1} \ldots \delta^{i_pj_p}\, \sum_{\underline{k}\neq\underline{0}}\, \sum_{\substack{\alpha_1=1 \\ \ldots \\ \alpha_{p-1} = 1}}^{d-1} \varepsilon_{j_1\cdots\, j_p}^{\alpha_1\cdots\alpha_p}(\underline{k})\, E^{\alpha_1\cdots\alpha_p}(\underline{k})\, e^{2\, i\, \pi\, \underline{k}(\vec{x}) } \, ,
\end{displaymath}
where $E^{\alpha_1\cdots\alpha_p}(\underline{k})$ is a complex valued antisymmetric tensor and $\underline{k}$ are discrete vectors of the torus dual lattice of which the components are measured in units of $L^{-1}$. Their norm is expressed as $\omega(\underline{k}) = \sqrt{k_i\, k_j\, \delta^{ij}}$. Note that the zero modes of the fields are not included in these expressions, as emphasized by the subscript $\bot$. In fact, these zero modes are the global degrees of freedom which have already been dealt with in the previous Section. The real valued tensors $\varepsilon^{\alpha_1 \cdots \alpha_p}_{i_1 \cdots i_p}(\underline{k})$ define a basis of orthonormalised polarisation tensors for each $\underline{k}\neq\underline{0}$. In our conventions, these tensors are constructed from a orthonormalised basis of polarisation vectors $\varepsilon^\alpha_i(\underline{k})$ for a vector field such that
\begin{equation}
   \varepsilon^\alpha_i(\underline{k})\, \varepsilon^\beta_j(\underline{k})\, \delta^{ij} = \delta^{\alpha\beta} ,
\end{equation}
where $\delta^{\alpha\beta}$ is the Kronecker symbol in polarisation space. This basis is chosen in such a way that, for each $\underline{k}\neq\underline{0}$, the dual lattice vector $\underline{\varepsilon}^d(\underline{k})$ is longitudinal whereas the vectors $\underline{\varepsilon}^\alpha(\underline{k})$ are transverse for $\alpha=1,\cdots,d-1$. Finally, it is convenient to choose for the longitudinal vector
\begin{displaymath}
   \underline{\varepsilon}^d(\underline{k}) = \frac{\underline{k}}{\omega(\underline{k})} , \qquad
\underline{k}\ne \underline{0} .
\end{displaymath}
Given the recursion relation induced by the Hodge decomposition theorem, the general polarisation tensor of any $p$-tensor field may be expressed as
\begin{displaymath}
   \varepsilon_{i_1\cdots i_p}^{\alpha_1\cdots\alpha_p}(\underline{k}) = \frac{1}{p!}\, \varepsilon_{\left[ i_1 \right.}^{\, \alpha_1}(\underline{k}) \ldots \varepsilon_{\left. i_p \right]}^{\, \alpha_p}(\underline{k}),
\end{displaymath}
which may likewise be decomposed into transverse and longitudinal components,
\begin{equation} \label{def:dyn_pol}
   \textrm{Longitudinal : } \left\lbrace \varepsilon_{i_1\cdots\,\, i_{p-1}\, i_p}^{\alpha_1\cdots\alpha_{p-1}\, d}(\underline{k}) \right\rbrace_{\alpha_1,\ldots,\alpha_{p-1}=1}^{d-1} ; \qquad
   \textrm{Transverse : } \left\lbrace \varepsilon_{i_1\cdots i_p}^{\alpha_1\cdots\alpha_p}(\underline{k}) \right\rbrace_{\alpha_1,\ldots,\alpha_p=1}^{d-1} \, .
\end{equation}
Given any mode, the $C^p_d$ degrees of freedom of a phase space field then separate into $C^{p-1}_{d-1}$ longitudinal and $C^{p}_{d-1}$ transverse degrees of freedom. These notations having been specified, and using the decompositions defined in (\ref{def:Hodge_normal}), the relevant quantum operators are Fourier expanded as
\begin{eqnarray}
\hat{E}_\bot^{i_1 \cdots i_p}(\vec{x}\,) & = & 
   \sum_{\underline{k} \neq \underline{0}} 
      \left\lbrace \delta^{i_1 j_1} \ldots \delta^{i_p j_p}\, p\, \sum_{\substack{\alpha_1=1 \\ \ldots \\ \alpha_{p-1} = 1}}^{d-1} \varepsilon_{j_1\cdots\,\, j_{p-1}\, j_p}^{\alpha_1\cdots\alpha_{p-1}\, d}(\underline{k})\, \hat{E}_L^{\alpha_1\cdots\alpha_{p-1}}(\underline{k}) \right.   \nonumber\\
   & + & 
      \left. \kappa\, \frac{\epsilon^{i_1\cdots i_p\, j_1\cdots j_{d-p}}}{(d-p-1)!} \sum_{\substack{\alpha_1=1 \\ \ldots \\ \alpha_{d-p-1} = 1}}^{d-1} \varepsilon_{j_1\cdots\,\, j_{d-p-1}\, j_{d-p}}^{\alpha_1\cdots\alpha_{d-p-1}\, d}(\underline{k})\, \hat{Q}_G^{\alpha_1\cdots\alpha_{d-p-1}}(\underline{k}) \right\rbrace\, e^{2\, i\, \pi\, \underline{k}(\vec{x})} , \nonumber\\
\hat{G}_\bot^{i_p \cdots i_{d-p}}(\vec{x}\,) & = & 
   \sum_{\underline{k} \neq \underline{0}} 
      \left\lbrace \delta^{i_1 j_1} \ldots \delta^{i_{d-p} j_{d-p}}\, (d-p) \sum_{\substack{\alpha_1=1 \\ \ldots \\ \alpha_{d-p-1} = 1}}^{d-1} \varepsilon_{j_1\cdots\,\, j_{d-p-1}\, j_{d-p}}^{\alpha_1\cdots\alpha_{d-p-1}\, d}(\underline{k})\, \hat{G}_L^{\alpha_1\cdots\alpha_{d-p-1}}(\underline{k}) \right. \nonumber\\ 
   & + & 
      \left. \frac{\kappa}{(p-1)!}\, \epsilon^{j_1\cdots j_p\, i_1\cdots i_{d-p}} \sum_{\substack{\alpha_1=1 \\ \ldots \\ \alpha_{p-1} = 1}}^{d-1} \varepsilon_{j_1\cdots\,\, j_{p-1}\, j_p}^{\alpha_1\cdots\alpha_{p-1}\, d}(\underline{k})\, \hat{P}_E^{\alpha_1\cdots\alpha_{p-1}}(\underline{k}) \right\rbrace\, e^{2\, i\, \pi\, \underline{k}(\vec{x})} \nonumber \, .
\end{eqnarray}
The self-adjoint property of the operator $\hat{E}^{i_1 \cdots i_p}(\vec{x}\,)$ translates into the following relations between the associated mode operators and their adjoint,
\begin{eqnarray}
\sum_{\substack{\alpha_1=1 \\ \ldots \\ \alpha_{p-1} = 1}}^{d-1} \varepsilon_{j_1\cdots\,\, j_{p-1}\, j_p}^{\alpha_1\cdots\alpha_{p-1}\, d}(\underline{k})\, \hat{E}_L^{\alpha_1\cdots\alpha_{p-1}}(\underline{k}) & = & \sum_{\substack{\alpha_1=1 \\ \ldots \\ \alpha_{p-1} = 1}}^{d-1} \varepsilon_{j_1\cdots\,\, j_{p-1}\, j_p}^{\alpha_1\cdots\alpha_{p-1}\, d}(-\underline{k})\, \hat{E}_L^{\dagger \,\, \alpha_1 \cdots \alpha_{p-1}}(-\underline{k}) , \nonumber\\
\sum_{\substack{\alpha_1=1 \\ \ldots \\ \alpha_{d-p-1} = 1}}^{d-1} \varepsilon_{i_1\cdots\,\, i_{d-p-1}\, i_{d-p}}^{\alpha_1\cdots\alpha_{d-p-1}\, d}(\underline{k})\, \hat{Q}_G^{\alpha_1\cdots\alpha_{d-p-1}}(\underline{k}) & = &\sum_{\substack{\alpha_1=1 \\ \ldots \\ \alpha_{d-p-1} = 1}}^{d-1} \varepsilon_{i_1\cdots\,\, i_{d-p-1}\, i_{d-p}}^{\alpha_1\cdots\alpha_{d-p-1}\, d}(-\underline{k})\, \hat{Q}_G^{\dagger \,\, \alpha_1 \cdots \alpha_{d-p-1}}(-\underline{k}) . \nonumber
\end{eqnarray}
Similar relations apply for the modes of the self-adjoint operator $\hat{G}^{i_p \cdots i_{d-p}}(\vec{x}\,)$.

Consequently, this decomposition of the non zero modes of the field operators in the dynamical sector leads to two decoupled subsectors, each of which is comprised of a countable set of mode operators with $\underline{k}\neq\underline{0}$. In the first subsector one has the operators $\hat{E}_L (\underline{k})$ and $\hat{P}_E (\underline{k})$ with the following non vanishing commutation relations,
\begin{equation}\label{eq:dyn_sec1_brackets}
   \left[ \hat{E}_L^{\dagger \,\, \alpha_1 \cdots \alpha_{p-1}}(\underline{k}) , \hat{P}_E^{\beta_1\cdots\beta_{p-1}}(\underline{k}') \right] = i\, \frac{\hbar}{V}\, \delta^{\alpha_1\left[\beta_1\right.} \ldots \delta^{\alpha_{p-1}\left.\beta_{p-1}\right]} \delta_{\underline{k}\underline{k}'}  ,
\end{equation}
while in the second subsector the operators $\hat{G}_L (\underline{k})$ and $\hat{Q}_G (\underline{k})$ possess the commutator algebra,
\begin{equation}\label{eq:dyn_sec2_brackets}
   \left[ \hat{G}_L^{\dagger \,\, \alpha_1 \cdots \alpha_{d-p-1}}(\underline{k}) , \hat{Q}_G^{\beta_1\cdots\beta_{d-p-1}}(\underline{k}') \right] = i\, \frac{\hbar}{V}\, \delta^{\alpha_1\left[\beta_1\right.} \ldots \delta^{\alpha_{d-p-1}\left.\beta_{d-p-1}\right]} \delta_{\underline{k}\underline{k}'}  ,
\end{equation}
$V$ being the volume of the space torus $\Sigma=T_d$.

This Fourier mode decomposition reduces the problem of diagonalising the Hamiltonian to a simple exercise in decoupled quantum oscillators, with
\begin{eqnarray}
\hat{H}_1[\hat{E}_L,\hat{P}_E] & = & \frac{V}{2\, (p-1)!}\, \sum_{\underline{k}\neq\underline{0}} \left\lbrace
        \kappa^2\, g^2\, \left( \hat{P}_E^{\alpha_1 \cdots \alpha_{p-1}}(\underline{k}) \right)^2 
        + \frac{\tilde{\omega}^2(\underline{k})}{\kappa^2\, g^2}\, \left( \hat{E}_L^{\alpha_1\cdots \alpha_{p-1}}(\underline{k}) \right)^2 \right\rbrace , \label{def:dyn_H1}\\
\hat{H}_2[\hat{G}_L,\hat{Q}_E] & = & \frac{V\, \kappa^2\, e^2}{2\, (d-p-1)!}\, \sum_{\underline{k}\neq\underline{0}} \left\lbrace \left( \hat{Q}_G^{\alpha_1 \cdots \alpha_{d-p-1}}(\underline{k}) \right)^2 + \frac{\tilde{\omega}^2(\underline{k})}{\kappa^4\, e^4}\, \left( \hat{G}_L^{\alpha_1 \cdots \alpha_{d-p-1}}(\underline{k}) \right)^2 \right\rbrace . \label{def:dyn_H2}
\end{eqnarray}
In these expressions the following notation is being used,
\begin{displaymath}
   \left( \hat{E}_L^{\alpha_1 \cdots \alpha_{p-1}}(\underline{k}) \right)^2 = \sum^{d-1}_{\substack{\alpha_1,\ldots,\alpha_{p-1} = 1 \\ \beta_1,\ldots,\beta_{p-1} = 1}} \hat{E}_L^{\alpha_1 \cdots \alpha_{p-1}}(\underline{k})\, \hat{E}_L^{\dagger \,\, \beta_1 \cdots \beta_{p-1}}(\underline{k})\,\, \delta^{\alpha_1\beta_1} \ldots \delta^{\alpha_{p-1}\beta_{p-1}} \, .
\end{displaymath}
The operators (\ref{def:dyn_H1}) and (\ref{def:dyn_H2}) are nothing other than the Hamiltonians of a collection of $C^{p-1}_{d-1}$ and $C^{p}_{d-1}$ independent harmonic oscillators, respectively, all of angular frequency
\begin{displaymath}
   \tilde{\omega}(\underline{k}) = \sqrt{4\, \pi^2\, \omega^2(\underline{k}) + \mu^2} , \qquad \mu=e\,g\,\kappa .
\end{displaymath}
The physical spectrum may easily be constructed by introducing annihilation and creation operators associated to the algebras (\ref{eq:dyn_sec1_brackets}) and (\ref{eq:dyn_sec2_brackets}). The annihilation operators are defined by
\begin{eqnarray}
   \mathfrak{a}^{\alpha_1 \cdots \alpha_{p-1}}(\underline{k}) & = & \frac{1}{\kappa\, g}\, \sqrt{\frac{V\,\tilde{\omega}(\underline{k})}{2\, \hbar}}\, \left( \hat{E}_L^{\alpha_1 \cdots \alpha_{p-1}}(\underline{k}) + i\, \frac{g^2\,\kappa^2}{\tilde{\omega}(\underline{k})}\, \hat{P}_E^{\alpha_1 \cdots \alpha_{p-1}}(\underline{k}) \right) , \nonumber\\
   \mathfrak{b}^{\alpha_1 \cdots \alpha_{d-p-1}}(\underline{k}) & = & \frac{1}{\kappa\, e}\, \sqrt{\frac{V\,\tilde{\omega}(\underline{k})}{2\, \hbar}}\, \left( \hat{G}_L^{\alpha_1 \cdots \alpha_{d-p-1}}(\underline{k}) + i\, \frac{\kappa^2\, e^2}{\tilde{\omega}(\underline{k})}\, \hat{Q}_G^{\alpha_1 \cdots \alpha_{d-p-1}}(\underline{k}) \right) \nonumber ,
\end{eqnarray}
whereas the creation operators $\mathfrak{a}^{\dagger\,\,\alpha_1\cdots \alpha_{p-1}}(\underline{k})$ and $\mathfrak{b}^{\dagger\,\, \alpha_1\cdots \alpha_{d-p-1}}(\underline{k})$ are merely the adjoint operators of $\mathfrak{a}^{\alpha_1 \cdots \alpha_{p-1}}(\underline{k})$ and $\mathfrak{b}^{\alpha_1 \cdots \alpha_{d-p-1}}(\underline{k})$, respectively. One then establishes the Fock algebras,
\begin{eqnarray}\label{eq:dyn12_Fock}
\left[\mathfrak{a}^{\alpha_1 \cdots \alpha_{p-1}}(\underline{k}),\mathfrak{a}^{\dagger\,\,\beta_1\cdots \beta_{p-1}}(\underline{k}')\right] & = & \delta^{\alpha_1\, \left[\beta_1\right.} \ldots \delta^{\alpha_{p-1}\, \left.\beta_{p-1}\right]}\, \delta_{\underline{k}\underline{k}'}, \nonumber \\
\left[\mathfrak{b}^{\alpha_1 \cdots \alpha_{d-p-1}}(\underline{k}),\mathfrak{b}^{\dagger\,\,\beta_1\cdots \beta_{d-p-1}}(\underline{k}')\right] & = & \delta^{\alpha_1\, \left[\beta_1\right.} \ldots \delta^{\alpha_{d-p-1}\, \left.\beta_{d-p-1}\right]}\, \delta_{\underline{k}\underline{k}'} ,
\end{eqnarray}
whereas (\ref{def:dyn_H1}) and (\ref{def:dyn_H2}) then reduce to the simple expressions,
\begin{eqnarray}
   \hat{H}_1[\mathfrak{a},\mathfrak{a}^\dagger] & = & \hbar\, \sum_{\underline{k}\ne\underline{0}} \tilde{\omega}(\underline{k}) \left( \frac{1}{2}\, C^{p-1}_{d-1} + \sum^{d-1}_{\alpha_1 < \cdots < \alpha_{p-1}} \mathfrak{a}^{\dagger\,\,\alpha_1\cdots \alpha_{p-1}}(\underline{k})\, \mathfrak{a}^{\alpha_1 \cdots \alpha_{p-1}}(\underline{k}) \right) , \nonumber\\
   \hat{H}_2[\mathfrak{b},\mathfrak{b}^\dagger] & = & \hbar\, \sum_{\underline{k}\ne\underline{0}} \tilde{\omega}(\underline{k}) \left( \frac{1}{2} \, C^{p}_{d-1} + \sum^{d-1}_{\alpha_1 < \cdots < \alpha_{d-p-1}} \mathfrak{b}^{\dagger\,\,\alpha_1\cdots \alpha_{d-p-1}}(\underline{k})\, \mathfrak{b}^{\alpha_1 \cdots \alpha_{d-p-1}}(\underline{k}) \right) \nonumber \, .
\end{eqnarray}
The Fock space representation is based on the normalised Fock vacuum $|0\rangle$, $\langle 0|0\rangle=1$, which is the kernel of all annihilation operators
\begin{displaymath}
 \mathfrak{a}^{\alpha_1 \cdots \alpha_{p-1}}(\underline{k}) \left\vert 0 \right\rangle = 0, \qquad \mathfrak{b}^{\alpha_1 \cdots \alpha_{d-p-1}}(\underline{k}) \left\vert 0 \right\rangle = 0, \qquad \varepsilon^{1+2}_{(0)} = \frac{1}{2}\hbar\, C^p_d\, \sum_{\underline{k}\neq\underline{0}} \tilde{\omega}(\underline{k})  ,
\end{displaymath}
where $\varepsilon^{1+2}_{(0)}$ is the divergent total vacuum energy. Excited states are obtained through the action onto the Fock vacuum of all $C^p_d=C^{p-1}_{d-1}+C^{p}_{d-1}$ creation operators, see (\ref{eq:dyn12_Fock}). This leads to states $\left\vert n_\gamma(\underline{k}) \right\rangle$ with energy eigenvalues
\begin{equation}\label{eq:dyn12_Energy}
\varepsilon_{\left(n_\gamma(\underline{k})\right)} = \varepsilon^{1+2}_{(0)} + \hbar\, \sum_{\underline{k}\neq\underline{0}} \sum_{\gamma=1}^{C^p_d} n_\gamma(\underline{k})\, \tilde{\omega}(\underline{k})  ,
\end{equation}
where $\left\lbrace n_\gamma(\underline{k}) \right\rbrace_{\gamma=1}^{C_p^d}$ are positive integers corresponding to number operator eigenvalues. A shorthand notation is used in (\ref{eq:dyn12_Energy}) with the index $\gamma$ labelling the $C^p_d$ possible combinations of a set of $p$ distinct integers in the range $[1,d]$, $\left\lbrace \alpha_1,\ldots,\alpha_i,\ldots,\alpha_p \right\rbrace_{\alpha_i=1}^{d}$, which will be referred to as $\Gamma_d^p$.

\section{Spectrum and Projection onto the TFT Sector}
\label{sec:Projection}

\subsection{Physical spectrum on the torus}

Combining all the results of the previous Sections for what concerns the diagonalisation of the physical TMGT Hamiltonian on the spatial $d$-torus $\Sigma=T_d$, the complete energy spectrum of states is given as
\begin{equation} \label{eq:Total_Energy}
\varepsilon_{\left(n_\gamma(\underline{k})\right)} = \varepsilon_{(0)} + \hbar\, \sum_{\underline{k}\in\mathbb{Z}^d} \sum_{\gamma} n_\gamma(\underline{k})\, \tilde{\omega}(\underline{k}) \, ,
\end{equation}
which is the sum of the contributions (\ref{eq:Global_dyn_Energy}) and (\ref{eq:dyn12_Energy}). Note that on the $d$-torus, the $p^{\textrm{th}}$ Betti number, $N_p$, equals $C^p_d$. The components of the vector $\underline{k}$ of the dual lattice may take any integer values since it is implicit in (\ref{eq:Total_Energy}) that $\left\lbrace n_\gamma(\underline{0})\right\rbrace_{\gamma=1}^{C^p_d} = \left\lbrace n_\gamma\right\rbrace _{\gamma=1}^{N_p}$. However, the index $\gamma$ has a different meaning whether $\underline{k}\neq\underline{0}$ or $\underline{k}=\underline{0}$. In the first case it refers to a value in the set $\Gamma_d^p$ and denotes one of the possible $C^p_d$ polarisations, while in the second case it is a (co)homology index, $\gamma=1,\cdots,{C_d^p}$. The total vacuum energy $\varepsilon_{(0)}$ in (\ref{eq:Total_Energy}) is divergent,
\begin{displaymath}
\varepsilon_{(0)} = \frac{1}{2}\hbar\, C^p_d\, \sum_{\underline{k}\in\mathbb{Z}^d} \tilde{\omega}(\underline{k}) \, ,
\end{displaymath}
and must be subtracted from the energy spectrum.

The positive integer valued functions $n_\gamma(\underline{k})$ count, for each $\underline{k} \neq \underline{0}$, the number of massive quanta of a $p$- or $(d-p)$-tensor field of momentum $2\pi\hbar\,\underline{k}$, of polarisation (\ref{def:dyn_pol}), namely
\begin{displaymath}
   \textrm{Transverse : } \varepsilon_{i_1\cdots i_p}^{\gamma} (\underline{k}), \quad \gamma \in \Gamma^p_{d-1}; \qquad \textrm{Longitudinal : } \varepsilon^{\gamma\, d}_{i_1\cdots i_p} (\underline{k}) , \quad \gamma \in \Gamma^{p-1}_{d-1} ,
\end{displaymath}
and of rest mass\footnote{A quantity indeed positive under the assumptions made.}
\begin{equation}\label{def:Mass_gap}
   M = \hbar\, \mu = \hbar\, \kappa\, e\, g  .
\end{equation}
There are also the contributions of the global quanta of the $p$- and $(d-p)$-tensor fields, where $\left\lbrace n_\gamma(\underline{0})\right\rbrace_{\gamma=1}^{C^p_d}$ count the numbers of excitations along the homology cycle generators $\Sigma_{(p)}^{\gamma}$ and $\Sigma_{(d-p)}^{\gamma}$. In the particular case when $p=1$, the integers $\left\lbrace n_\gamma(\underline{k}) \right\rbrace_{\gamma=1}^{d}$ count, for each $\underline{k} \neq\underline{0}$, the number of massive photons of momentum $2\pi\hbar\,\underline{k}$, of rest mass $M$ and of polarisation
\begin{displaymath}
\textrm{Transverse :}\ \left\lbrace \varepsilon_i^\gamma (\underline{k}) \right\rbrace_{\gamma=1}^{d-1}, \qquad 
\textrm{Longitudinal :}\ \varepsilon^d_i (\underline{k}) \, .
\end{displaymath}

Depending on how one deals with large gauge transformations in the TFT sector, each energy state is either infinitely degenerate for a real valued $k$, see (\ref{eq:kappa_quant}), or ($\prod_{\delta=1}^{N_p} k_1\, k_2\, \mathcal{I}\, \textrm{Min}(I_{\delta\delta'})$) times degenerate if $k$ is a rational number of the form $k=k_1 / k_2$. If $k$ is an integer, each energy state is ($\prod_{\delta=1}^{N_p} k\, \mathcal{I}\, \textrm{Min}(I_{\delta\delta'})$) times degenerate and the mass gap is then quantised,
\begin{displaymath}
   M = \frac{\hbar^2}{2\pi}\, \mathcal{I}\, k\, e\, g \, .
\end{displaymath}
In the Maxwell-Chern-Simons (MCS) case in $2+1$ dimensions, we recover in the global sector a quantum mechanical system corresponding to the Landau problem of condensed matter physics on the 2-torus.

\subsection{Projection onto the topological field theory sector}

At least formally, the naive limit of infinite coupling constants, $e\rightarrow\infty$ and $g\rightarrow\infty$, in the classical Lagrangian of topologically massive gauge theories, see (\ref{def:TMGT_Action}) and (\ref{def:MCS_Action}), must lead to a pure topological field theory (TFT) of the $B\wedge F$ or $A\wedge F$ type, see (\ref{def:BF_Action}) or (\ref{def:AF_Action}). However, as pointed out by several authors (see for example \cite{Polychronakos:1989cd,Dunne:1989hv}), a paradox seems to arise at the quantum level (as well as within the classical Hamiltonian formulation) when the pure Chern-Simons (CS) theory is viewed as the limit $e \to \infty$ of the Maxwell-Chern-Simons (MCS) theory. The Hilbert space of the CS theory is constructed from the algebra of the non commuting configuration space operators which are in fact canonically conjugate phase space operators. As far as the MCS theory is concerned, its Hilbert space is constructed from the Heisenberg algebras of twice as many phase space operators. This problem is generic whenever a pure quantum TFT (TQFT) is considered as the limit of its associated TMGT because two distinct Hilbert spaces are being compared. Actually, due to the second-class constraints appearing in the Hamiltonian analysis of a TFT which is already in Hamiltonian form, non vanishing commutation relations apply to the configuration space operators. Furthermore, the Gauss law constraints of pure TFT are not the limit of the Gauss law constraints of TMGT. The former operators tend to restrict too drastically the physical Hilbert space in comparison to the limit of the TMGT physical Hilbert space.

This problem of an ill-defined limit is usually handled by projecting from the Hilbert space of the TMGT onto its degenerate ground state. This projection actually acts in a manner similar to second-class constraints which then lead to a reduced phase space and non vanishing configuration space commutation relations determined from the associated Dirac brackets. The global sector of the MCS theory is analogous to the classical Landau problem of a charged point particle of mass $m$ moving in a two dimensional surface in the presence of an uniform external magnetic field $B$ perpendicular to that surface. Within the latter context the mass gap (\ref{def:Mass_gap}) corresponds to the cyclotron frequency $\omega_c$ \cite{Dunne:1989hv,Dunne:1998qy}. The spectrum of the quantised model is organised into Landau levels (with a degeneracy dependent on the underlying manifold), of which the energy separation $\omega_c$ is proportional to the ratio $B/m$. The limit $B\to\infty$ or $m\to 0$ effectively projects onto the lowest Landau level (LLL) in which one obtains a non commuting algebra for the space coordinates. By analogy, projection onto the ground state reduces the phase space of the MCS theory to the canonically conjugate configuration space operators of a pure CS theory. In the global sector, the projection from a general TMGT, (\ref{def:TMGT_Action}), to a pure TQFT offers in some sense a generalisation of the LLL projection in any dimension. Considering that the mass gap (\ref{def:Mass_gap}) of the TMGT becomes infinite for coupling constants running to infinity, all excited states decouple from the physical spectrum, leaving over only the degenerate ground states. Projection onto these ground states restricts the Hilbert space to that of a TQFT.

Interestingly, the PT factorisation established in this work enables the usual projection from TMGT to TQFT to be defined in a natural way. Already in the classical Hamiltonian formulation phase space is separated into two decoupled sectors, the first being dynamical and manifestly gauge invariant, and the second being equivalent to a pure TFT with identical Gauss law constraints and commutation relations. As a matter of fact in the present approach which does not require any gauge fixing procedure whatsoever, the non commuting sector of a CS theory or, more generally, the reduced phase space of a TQFT appears no longer after the projection onto the ground state at the quantum level (or after the introduction of Dirac brackets) but is manifest already at the classical Hamiltonian level. By letting $e$ or $g$ grow infinite, the mass gap (\ref{def:Mass_gap}) becomes infinite, hence dynamical massive excitations decouple whereas the TFT sector, which is independent of the coupling constants, remains unaffected. In this limit, the system looses any dynamics, the latter being intimately related to the Riemannian metric structure of the spacetime manifold, while all that is then left is a wave function depending on global variables only, namely the quantum states of a TQFT.

\section{Conclusion and Outlook}

The main result of this paper is the identification of a Physical-Topological (PT) factorisation of the classical phase space of abelian topologically massive gauge theories (TMGT) in any dimension, into a manifestly gauge invariant and dynamical sector of non commuting ``electric fields" and a gauge variant purely topological sector of the $B\wedge F$ or $A\wedge F$ type. This factorisation is achieved through a canonical transformation in the phase space of TMGT. The discussion considers the most general action for abelian TMGT in any dimension and for any $p$-form fields, including the two possible types of topological terms related through an integration by parts. The clue to this PT factorisation relies on the identification of a topological field theory embedded in the full TMGT, which is not manifest in the original Hamiltonian formulation. Let us emphasize that the procedure does not require any gauge fixing choice whatsoever, with its cortege of second-class constraints or ghost degrees of freedom. Rather, the PT classical factorisation readily allows for a straightforward quantisation of these systems and the identification of their spectrum of gauge invariant physical states, accounting also for all the topological features inherent to such dynamics.

In the early 1990's, A.~P.~ Balachandran and P.~Teotonio-Sobrinho \cite{Balachandran:1992qg} established that it is possible to identify among the phase space variables of a TMGT combinations corresponding to those of a TFT. They noticed that, in a particular case of an underlying manifold with boundary, edge states may be understood in terms of a TFT, already at the classical level. Nevertheless, this paper did not realise the powerful gauge fixing free factorisation of the theory into two decoupled sectors as described in the present work. Incidentally, it should be of interest to analyse how this new approach may shed new light onto this paper, in a manner akin to that in which it makes most transparent and natural the projection onto a topological field theory through the limits $e, g\to\infty$, generalising the concept of projection onto the lowest Landau level of the Landau problem. In the present approach, the TFT sector with its reduced phase space is actually made manifest already at the classical level, independently of any projection onto the quantum ground state, or any classical projection onto physical edge states in the case of a manifold with boundary.

The formalism of TMGT defined by the actions in (\ref{def:TMGT_Action}) or (\ref{def:MCS_Action}) offers a possible description of some phenomenological phenomena such as effective superconductivity \cite{Balachandran:1992qg,Diamantini:1995yb}, Josephson arrays~\cite{Diamantini:2005dj}, etc. Furthermore, it is certainly of interest to investigate the perspectives offered by the PT factorisation when a $B\wedge F$ or $A\wedge F$ field theory is coupled to (non)relativistic matter fields as an effective description of phenomena related, for example, to QCD confinement \cite{Chatterjee:2006iq}. It is also of interest to extend this approach to Yang-Mills-Chern-Simons theories \cite{MCS,Dunne:1998qy} or to the nonabelian generalisation of the Cremmer-Scherk theory which requires the introduction of extra fields or to allow for non renormalisable couplings since the generalisation to a local, power counting renormalisable action while preserving the same field content and the same number of local symmetries as the abelian theory (\ref{def:TMGT_Action}) is not possible (see \cite{Henneaux:1997mf} and references therein). However the long term goal is to gain a deeper understanding of the influence of topological terms, such as the topological mass gap, and of topological sectors in field configuration space on the nonperturbative dynamics of gauge theories, beginning with Yang-Mills theories coupled to whether fermionic or bosonic matter fields.

\section*{Acknowledgments}

The work of B.~B. is supported by a Ph.D. Fellowship of
the ``Fonds pour la formation \`a la Recherche dans l'Industrie et dans l'Agriculture'' (FRIA),
of the Associated Funds of the National Fund for Scientific Research (F.N.R.S., Belgium).

J.~G. acknowledges the Institute of Theoretical Physics for an Invited Research Staff position at the
University of Stellenbosch (Republic of South Africa).
He is most grateful to Profs. Hendrik Geyer and Frederik Scholtz, and the School of Physics
for their warm and generous hospitality during his sabbatical leave, and for financial support.
His stay in South Africa is also supported in part by the Belgian National
Fund for Scientific Research (F.N.R.S.) through a travel grant.

J.~G. acknowledges the Abdus Salam International Centre for Theoretical
Physics (ICTP, Trieste, Italy) Visiting Scholar Programme
in support of a Visiting Professorship at the ICMPA.

This work is also supported by the Institut Interuniversitaire des Sciences Nucl\'eaires and by
the Belgian Federal Office for Scientific, Technical and Cultural Affairs through
the Interuniversity Attraction Poles (IAP) P5/27 and P6/11.

\end{document}